\documentclass{webofc}
\usepackage[varg]{txfonts}

\usepackage{graphicx}
\usepackage{subcaption}

\newcommand{\cD}{{\cal D}}

\newcommand{\cA}{{\cal A}}

\newcommand{\cP}{{\cal P}}
\newcommand{\be}{\begin{equation}}
\newcommand{\ee}{\end{equation}}

\newcommand{\rkx}{\right)}
\newcommand{\lk}{\left(}

\newcommand{\sli}{\sum\limits}

\newcommand{\il}{\int\limits}

\newcommand{\ve}{\vec{e}}

\newcommand{\vp}{\vec{p}}

\newcommand{\vz}{\vec{z}}
\newcommand{\vw}{\vec{w}}

\newcommand{\va}{\vec{a}}

\newcommand{\vA}{\vec{A}}
\newcommand{\vD}{\vec{D}}
\newcommand{\vx}{\vec{x}}

\newcommand{\vB}{\vec{B}}
\newcommand{\vy}{\vec{y}}
\newcommand{\vE}{\vec{E}}
\newcommand{\vq}{\vec{q}}

\newcommand{\ii}{\mathrm{i}}
\newcommand{\dd}{\mathrm{d}}

\renewcommand{\vec}[1]{\mbox{\boldmath$#1$\unboldmath}}

\newcommand{\R}{\mathbb{R}}

\newcommand*{\varpm}{\mathbin{\ooalign{\hfil$\pm$\hfil\cr\hfil\raise-.3ex\hbox{$\scriptscriptstyle(\mkern16mu)$}\hfil}}}

\woctitle{4\textsuperscript{th} International Conference on New Frontiers in Physics (ICNFP 2015)}
\begin{document}
\title{Hamiltonian approach to QCD in Coulomb gauge: From the vacuum to finite temperatures}

\author{H. Reinhardt\inst{1}\fnsep\thanks{\emph{speaker}}\fnsep\thanks{\email{hugo.reinhardt@uni-tuebingen.de}} \and
        D. Campagnari\inst{1}\fnsep\thanks{\email{d.campagnari@uni-tuebingen.de}} \and
        J. Heffner\inst{1}\fnsep\thanks{\email{heffner@tphys.physik.uni-tuebingen.de}} \and
             M. Quandt\inst{1}\fnsep\thanks{\email{markus.quandt@uni-tuebingen.de}} \and
             P. Vastag\inst{1}\fnsep\thanks{\email{peter.vastag@uni-tuebingen.de}}
}

\institute{Universit\"at T\"ubingen\\ Institut f\"ur Theoretische Physik\\ Auf der Morgenstelle 14\\
D-72076 T\"ubingen}

\abstract{%
 The variational Hamiltonian approach to QCD in Coulomb gauge is reviewed and the essential 
 results obtained in recent years are summarized. First the results for the vacuum sector are 
 discussed, with a special emphasis on the mechansim of confinement and chiral symmetry breaking. 
 Then the deconfinement phase transition is described by introducing temperature in the Hamiltonian 
 approach via compactification of one spatial dimension. The effective action for the 
 Polyakov loop is calculated and the order of the phase transition as well as the critical 
 temperatures are obtained for the color group SU(2) and SU(3). In both cases, our predictions 
 are in good agreement with lattice calculations.
}
\maketitle
\section{Introduction}
\label{intro}
\begin{figure}
\centering
\includegraphics[width=8cm,clip]{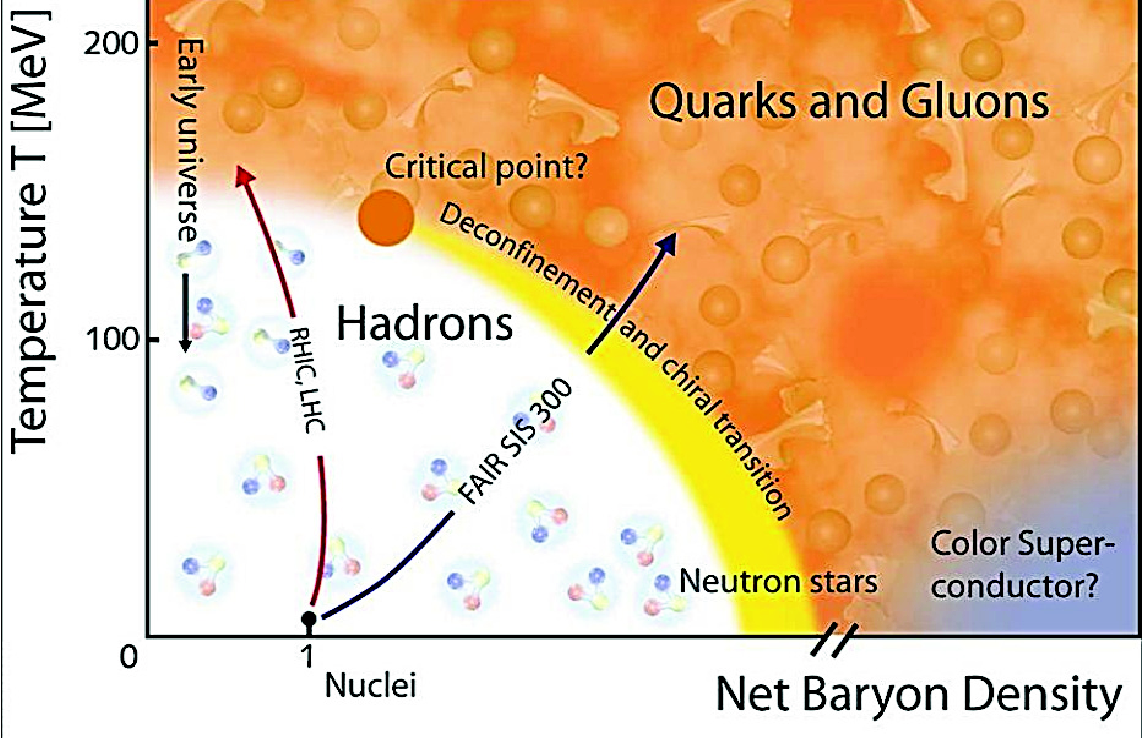}
\caption{\label{fig-1}The phase diagram of QCD.}
\end{figure}

Understanding the QCD vacuum is still one of the big challenges of theoretical physics. Although substantial 
progress has been made during the last two decades we are far  from rigorously understanding the essential 
features of the QCD vacuum: confinement and spontaneous breaking of chiral symmetry. Under normal conditions
quarks and gluons are confined inside hadrons, and the hadrons receive most of their mass through the mechanism
of spontaneous breaking of chiral symmetry. When baryonic matter is heated up or strongly compressed, hadrons
melt and a plasma of quarks and gluons is formed. This deconfinement phase transition is accompanied by 
the restoration of chiral symmetry. We have learned a lot about this phase transition from lattice calculations
at finite temperatures and vanishing baryon density, i.e.~along the temperature axis of the phase diagram, 
see Fig.~\ref{fig-1}. However, the lattice approach fails at finite chemical potential due to the so-called sign 
problem (the quark determinant becomes complex). To understand the deconfinement phase transition 
at finite chemical potential continuum approaches are thus required, which do not face the sign problem and 
are able to deal with a non-zero chemical potential. I will present here such a continuum formulation, namely the 
\emph{Hamiltonian approach}. In fact, this method is currently being developed and tested at finite temperatures 
where it can be compared to available lattice calculations, and I will not yet present results for finite 
chemical potential. In the first part of my talk I will study the QCD vacuum within 
this approach and will then extend it to finite temperatures.

One may ask for the benefit of the Hamiltonian approach as compared to the more common functional integral approach, 
which is so elegant for perturbation theory and which is also the basis for the lattice calculations. From quantum 
mechanics we know that solving the Schr\"odinger equation is usually much simpler and much more efficient than 
calculating the functional integral directly. For example, it is quite easy to solve the Schr\"odinger equation 
for the hydrogen atom, while calculating the corresponding functional integral directly is exceedingly complicated. 
Also in non-perturbative continuum quantum field theory the Hamiltonian approach is in my opinion more efficient. 
In my talk I will use the Hamiltonian approach to QCD in Coulomb gauge first to study the QCD vacuum by means of a 
variational solution of the Yang--Mills Schr\"odinger equation. Then I will extend this  approach to finite 
temperatures by compactifying a spatial dimension. I will first illustrate the method by means of free bosons and 
fermions and then apply it to QCD at finite temperatures, where I will mainly concentrate on the calculation 
of the effective potential of the Polyakov loop. 

\section{Hamiltonian approach to QCD in Coulomb gauge}
\label{sec-1}
The Hamiltonian approach to quantum field theory is based on the canonical quantization. We illustrate this 
formalism first for the Yang--Mills theory, which is defined by the classical action $S = \frac{1}{4}
-\int \dd^4 x \, F_{\mu \nu}^a F^{a \mu \nu}$. Here, $F^a_{\mu \nu} = \partial_\mu A^a_\nu - \partial_\nu A^a_\mu +
g f^{abc} A^b_\mu A^c_\nu$ denotes the non-Abelian field strength. For the canonical quantization, the gauge fields are considered as the 
``coordinates''. The canonical conjugate momenta are introduced in the standard fashion by $\Pi^a_\mu (\vx) = 
\delta S / \delta \dot{A}^{a \mu} (\vx)$. For the spatial components of the gauge field $A_i^a (\vx)$ these momenta coincide 
with the color electric field $\Pi^a_i (\vx) = - E^a_i (\vx)$, while the momentum corresponding to the temporal 
component of the gauge field $A_0^a (\vx)$ vanishes. The latter causes problems in the canonical quantization which, however,
can be circumvented by imposing Weyl gauge $A_0^a (\vx) = 0$. One finds then the following classical Hamiltonian of 
Yang--Mills theory
\be
\label{G1}
H = \frac{1}{2} \int \dd^3 x \lk \vec{\Pi}^a (\vx) \cdot \vec{\Pi}^a (\vx) + \vB^a (\vx) \cdot \vB^a (\vx) \rkx
\ee
where $B_i^a = \frac{1}{2} \varepsilon_{i j k} F_{j k}^a$ is the color magnetic field. The theory is quantized by imposing the usual canonical commutation relations, i.e.
\begin{align}
\label{G2}
\left[ A^a_k (\vx) , \Pi^b_l (\vy) \right] &= \ii \delta^{ab} \delta_{kl} \delta(\vx - \vy) \\
\left[ A^a_k (\vx) , A^b_l (\vy) \right] &= \left[ \Pi^a_k (\vx) , \Pi^b_l (\vy) \right] = 0 \ .
\end{align}
In the coordinate representation the momentum operator is therefore realized by 
$\Pi^a_k (\vx) = \delta / (\ii \delta A^a_k (\vx))$.

Because of the Weyl gauge $A_0^a (\vx) = 0$, Gau\ss's law is lost as an equation of motion and has to be 
imposed as a constraint on the wave functional
\be
\label{G3}
\hat{D}^{ab}_k (\vx) \Pi^b_k (\vx) \psi [A] = \rho_m^a (\vx) \psi [A] \, .
\ee
Here $ \hat{D}^{ab}_k (\vx) = \delta^{ab} \partial^x_k + g f^{acb} A^c_k (\vx)$
is the covariant derivative in the adjoint representation of the gauge group, and the operator on  
the left hand side, $\hat{\vD} \cdot \vec{\Pi}$, is nothing but the generator 
of time-independent (space-dependent) gauge transformations. Furthermore, $\rho_m$ is the color
charge density of the matter field. In the absence of a matter field the right hand side of eq.~(\ref{G3})
vanishes and the wave functional has to be invariant under time-independent gauge transformations $\psi [A^U] = \psi [A]$.
With the canonical momentum replaced by the operator $\Pi^a_k (\vx) = \delta / (\ii \delta A^a_k (\vx))$, eq.~(\ref{G1})
becomes the Hamilton operator of Yang--Mills theory. With this operator one has to solve then the Schr\"odinger equation
\be
\label{G4}
H \psi [A] = E \psi [A] \, 
\ee
in the Hilbert space of gauge invariant wave functionals, which has the usual scalar product known from 
quantum mechanics
\be
\label{G5}
\langle \phi | \ldots | \psi \rangle = \int \cD \vA \, \phi^* [\vA] \ldots \psi [\vA] \, ,
\ee
where $\int \cD \vA$ denotes the functional integral over (time-independent) spatial components of the gauge field. 
One is mainly interested in the solution of the Schr\"odinger equation (\ref{G4}) for the ground state (vacuum). 
There have been attempts
to solve the Schr\"odinger equation directly for the vacuum in the Hilbert space of gauge invariant wave functionals. 
However, so far these attempts have been successful only in 2+1 dimensions \cite{2}. In 3+1 dimensions it seems extremly difficult
to find gauge invariant representations of the wave functional except for some limited cases \cite{R1}. 
Therefore it is much more efficient to fix the gauge, and a convenient choice for this purpose is the Coulomb gauge 
$\vec{\partial} \cdot \vA = 0$.
This condition can be implemented in the scalar product (\ref{G5}) by the standard Faddeev-Popov method resulting in 
\be
\label{G6}
\langle \phi | \ldots | \psi \rangle = \int \cD \vA^\perp \, J (\vA^\perp) \phi^* [\vA^\perp] \ldots \psi [\vA^\perp] \, .
\ee
Here the integration is now over the transversal components of the gauge field $\vA^\perp$ only and 
\be
\label{G7}
J (\vA^\perp) = \mathrm{Det} (- \hat{\vD} \cdot \vec{\partial})
\ee
is the Faddeev-Popov determinant in Coulomb gauge. 
The Coulomb gauge fixing basically corresponds to a transition from Cartesian coordinates to curvilinear coordinates and 
the Faddeev-Popov determinant represents the Jacobian of this transformation.

Even if the longitudinal components of the gauge field are 
eliminated in Coulomb gauge the momentum operator still contains also a longitudinal component 
$\vec{\Pi}^{\parallel}$. From Gau\ss's law one finds for the longitudinal momentum operator when it acts on a wave functional 
\be
\label{G9}
\vec{\Pi}^{\parallel}\, \psi [\vA] = -\vec{\partial} ( - \hat{\vD} \cdot \vec{\partial})^{- 1} \rho\, \psi[\vA] \,, \quad \qquad
\rho^a = - f^{abc} A^b_k \Pi^c_k + \rho^a_m \, .
\ee
With this relation (\ref{G9}) one finds for the Yang--Mills Hamiltonian in Coulomb gauge \cite{3}
\begin{align}
 \label{G10}
 H &= \frac{1}{2} \int \dd^3 x \lk J^{- 1} \vec{\Pi}^a(\vx) \cdot J \vec{\Pi}^a(\vx) + \vB^a(\vx) \cdot \vB^a(\vx) \right) + H_C \nonumber\\
 & \equiv H_{YM} + H_C \, ,
 \end{align}
where 
\begin{align}
\label{G11}
 H_C = \frac{1}{2} \int \dd^3 x \int \dd^3 y \, J^{- 1} \rho^a(\vx) J \left[(- \hat{\vD} \cdot \vec{\partial} )^{- 1} 
 (- \vec{\partial}^2) (- \hat{\vD} \cdot \vec{\partial})^{- 1}\right]^{a b}(\vx, \vy) \rho^b(\vy)
\end{align}
arises from the kinetic energy of the longitudinal components of the momentum operator and is usually referred to 
as the Coulomb term. 
Here $\rho(\vx)$ is the total color charge density, which includes besides the matter charge density $\rho_m(\vx)$
also the charge density of the gauge field $\rho^a_g(\vx) = - f^{abc} A^b_k(\vx) \Pi^c_k(\vx)$, which exists obviously only in 
non-Abelian gauge theories.
In the case of QED where the Faddeev-Popov operator becomes the Laplacian 
$(- \Delta)$, the term eq.~(\ref{G11}) produces precisely the Coulomb potential 
between electric charge densities  $\rho$.

The Coulomb gauge fixed Hamiltonian (\ref{G10}) is obviously more complicated than the original gauge invariant one 
(\ref{G1}). First, the Faddeev-Popov determinant appears in the kinetic energy of the transversal degrees of freedom 
[the first term in eq.~(\ref{G10})] and, second, the kinetic energy of the longitudinal degrees of freedom (\ref{G11})
is a highly non-local object. Furthermore, the Faddeev-Popov determinant appears also in the scalar product (\ref{G6}).
This is the price one has to pay for the gauge fixing. Nevertheless it is still more convenient to work with 
this more complicated gauge fixed Hamiltonian than with gauge invariant wave functionals. Let us stress that by implementing 
Gau\ss{}'s law in the gauge fixed Hamiltonian, all consequences of gauge invariance are fully taken
into account. 

\section{Variational solution of the Yang--Mills Schr\"odinger equation for the vacuum}
\label{sec-2}
\begin{figure}
\centering
\begin{subfigure}{0.45\textwidth}
\includegraphics[width=\textwidth]{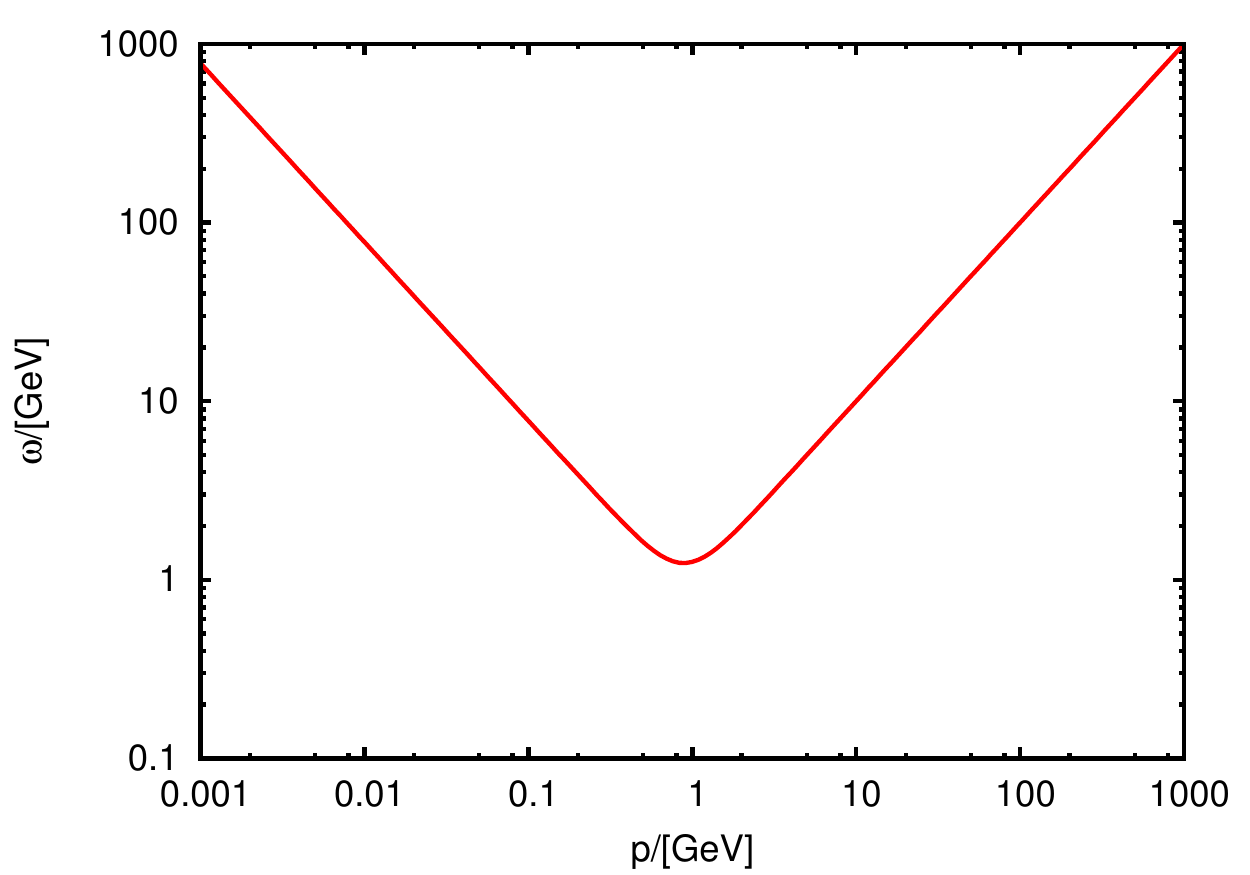}
\caption{}
\end{subfigure}
\quad
\begin{subfigure}{0.45\textwidth}
\includegraphics[width=\textwidth]{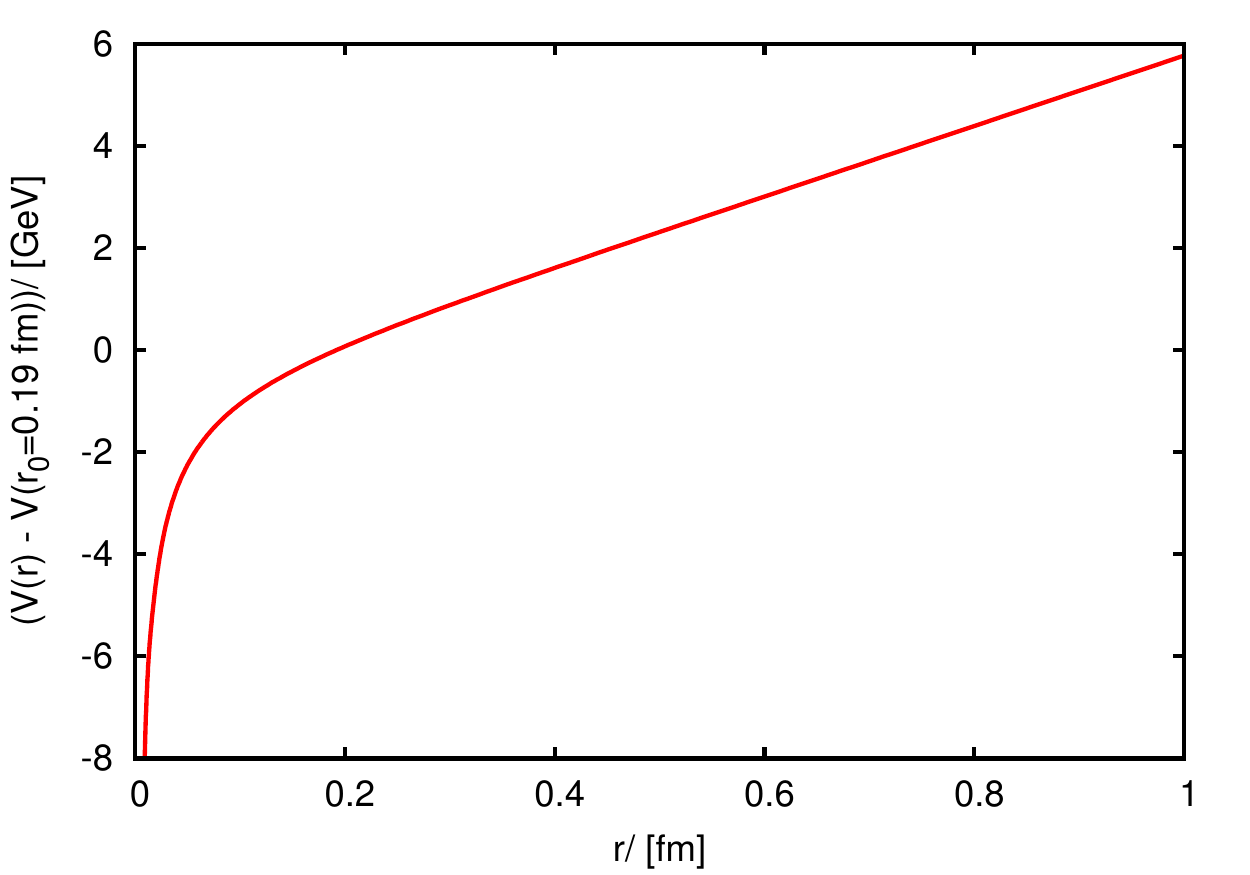}
\caption{}
\end{subfigure}
\caption{(a) Gluon energy $\omega (p)$ following from the variation of the Yang--Mills energy density. 
(b) Static color charge potential obtained in the variational approach \cite{7}. }
\label{fig-2}       % Give a unique label
\end{figure}

\begin{figure}
\centering
\includegraphics[width=2cm,clip]{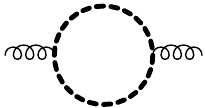}
\caption{Diagrammatic representation of the ghost loop.}
\label{fig-3}       % Give a unique label
\end{figure}

Variational calculations within the Hamiltonian approach to Yang--Mills theory in Coulomb gauge were 
initiated by D. Sch\"utte in Ref.~\cite{4} who used a Gaussian trial ansatz for the vacuum wave
functional. By minimizing the energy density he derived a closed  set of coupled integral equations 
for the gluon propagator, the ghost propagator and the Coulomb potential. Later on, Ref.~\cite{RX}
followed the approach of Ref.~\cite{4} by using the same ansatz for the vacuum wave functional, but 
in addition provided the first numerical solution of the resulting integral equations.
In Refs.~\cite{5,6} we have developed in our group a variational approach to Yang--Mills theory in Coulomb
gauge, which differs from previous approaches:
i) in the ansatz for the vacuum wave functional, ii) in the treatment of the Faddeev-Popov
determinant\footnote{In Refs.~\cite{4,RX} the Faddeev-Popov determinant was ignored which, however, 
turns out to be crucial for the infrared properties of the theory.}
and
iii) in the renormalization.
Our trial ansatz for the vacuum wave functional is given by 
\be
\label{G12}
\psi [A] = \frac{1}{\sqrt{J (A)}} \exp \left[ - \frac{1}{2} \int \dd^3 x \int \dd^3 y \, A^a_k (\vx)
\omega (\vx, \vy) A^a_k (\vy) \right] \, ,
\ee
where $\omega (\vx, \vy)$ is a variational kernel, which is determined by minimizing the energy density. 
The ansatz (\ref{G12}) has the advantage that the static gluon propagator is given by the inverse of the 
kernel $\omega$
\be
\label{G13}
\langle A^a_k (\vx) A^b_l (\vy) \rangle = \delta^{ab} t_{kl} (\vx) \frac{1}{2} \omega^{- 1} (\vx, \vy) \, ,
\ee
where $t_{kl} (\vx) = \delta_{kl} - \partial^x_k \partial^x_l / \partial^2_x$ is the transversal projector. 
From the form of the static gluon propagator (\ref{G13}) it follows that the Fourier transform of $\omega (\vx, \vy)$ 
represents the single-particle gluon energy. 
Minimizing the energy density with respect to $\omega (\vx, \vy)$ yields the result shown in Fig.~\ref{fig-2} (a) \cite{7}. For large momenta
the gluon energy behaves like the photon energy $\omega (p) \sim | \vp|$ while at small momenta it diverges $\omega (p)
\sim 1 / | \vp|$ implying the absence of free gluons in the infrared, which is a manifestation of confinement.
The numerical calculation carried out in Ref.~\cite{8} shows that the 
Coulomb term (\ref{G11}) can be ignored in the Yang--Mills sector. Then the gluon gap equation resulting from the 
minimization of the energy density is given by 
\be
\label{G14}
\omega^2 (p) = \vp^2 + \chi^2 (p) \, ,
\ee
where $\chi (p)$ is the ghost loop, which is diagrammatically represented in Fig.~\ref{fig-3}. The gap equation
(\ref{G14}) has the form of the dispersion relation of a relativistic particle except that the (effective)
mass of the particle is replaced by the ghost loop $\chi (p)$. 
The ghost propagator is given here as the vacuum expectation value of the inverse Faddeev-Popov operator 
and is represented by 
\be
\label{G25}
G = \langle (- \hat{\vD} \cdot \vec{\partial})^{- 1} \rangle = d (- \Delta) / (- \Delta) \, ,
\ee
where $d (- \Delta)$ is the so-called ghost form factor. Obviously, this form factor contains all the deviations from QED
where the Faddeev-Popov operator is just given by the Laplacian. For large momenta the ghost form factor approaches unity while 
it diverges like $1/p$  in the infrared and thus fulfills the so-called horizon condition 
$d^{- 1} (p = 0) = 0$ \cite{R2}. 
Contrary to other gauges, in Coulomb gauge the ghost form factor $d$ has a physical meaning: Its inverse represents
the dielectric function of the Yang--Mills vacuum, Ref.~\cite{9}. The horizon condition guarantees that the dielectric 
function vanishes in the infrared $\varepsilon (p = 0) = 0$, which implies that the Yang--Mills vacuum is a perfect color
dielectricum or a so-called dual superconductor. In this way the Hamiltonian approach in Coulomb gauge relates to the 
dual superconductor picture of confinement introduced many years ago by Mandelstam \cite{10} and 't Hooft \cite{11}. 
The form of the dielectric function obtained here is in accord with the phenomenological bag model picture of hadrons 
\cite{12}: On the one hand, inside the hadrons i.e.~at small distances, the dielectric function equals unity which is the 
value of a normal vacuum, as in QED and also in QCD without matter, due to asymptotic freedom.
On the other hand, outside the physical hadrons the infrared value $\varepsilon  (p = 0) = 0$
corresponding to a confining vacuum is realized. Indeed for a medium with $\varepsilon = 0$ the electrical 
displacement $\vD = \varepsilon \vE$ vanishes implying by Gau\ss{}'s law, $\vec{\partial} \cdot \vD = \rho_{free}$, 
that no free color charges can exist. 

\begin{figure}
\centering
\includegraphics[width=0.45\textwidth,clip]{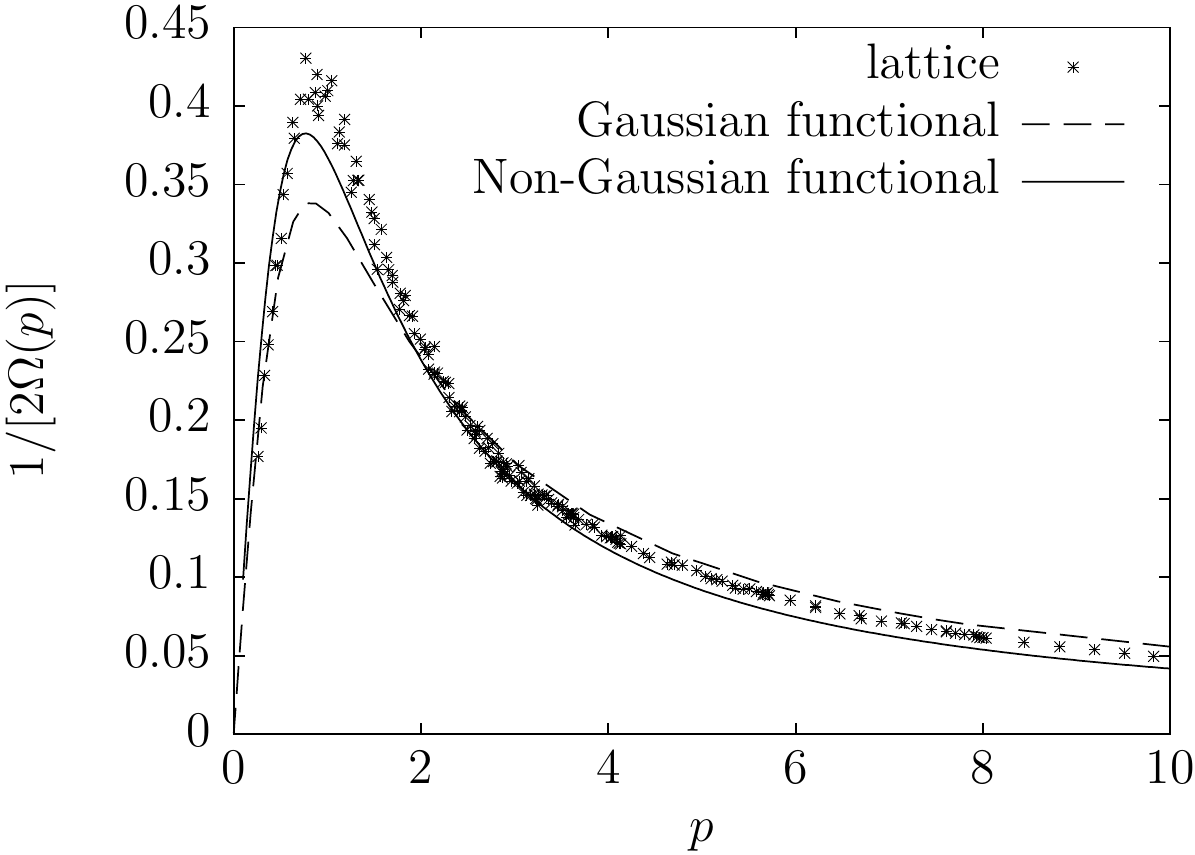}
\caption{The gluon propagator. Comparison of the lattice result (crosses) \cite{13} with the result of the variational calculation.}
\label{fig-4}       % Give a unique label
\end{figure}

Figure~\ref{fig-4} compares the results of the variational calculation with the lattice results for the gluon propagator. The 
lattice results \cite{13} can be nicely fitted by Gribovs' formula \cite{R3}
\be
\label{G26}
\omega (p) = \sqrt{\vp^2 + M^4 / \vp^2} 
\ee
with a mass scale of $M \simeq \, 880 \, \mbox{MeV}$. The gluon energy (dashed line) obtained with the Gaussian trial
wave functional agrees quite well with the lattice data in the infrared and in the ultraviolett regime but misses some 
strength in the mid-momentum regime. This missing strength is largely recovered when a non-Gaussian wave functional is 
used, Ref.~\cite{14}, which includes in the exponent of the trial wave functional also cubic  and quartic terms in the gauge 
potential. When non-Gaussian wave functionals are used, Wick's theorem is no longer valid. In Ref. \cite{14} this problem 
was circumvented by exploiting generalized Dyson-Schwinger equations to express the various $n$-point functions occuring in 
the expectation values of operators like the Hamiltonian in terms of the variational kernels occuring in the exponent
of the wave functional. One finds then the full curve in Fig.~\ref{fig-4}, which fits the lattice data in the 
mid-momentum regime much better than the Gaussian ansatz.

The part of the Coulomb Hamiltonian (\ref{G11}), which is quadratic in the charge density of the matter field $\rho_m$ 
represents a two-body interaction which is induced by the Yang--Mills vacuum. The Faddeev-Popov determinant 
drops out from this part of the Coulomb term and its vacuum expectation value yields the static color charge potential
\be
\label{G27}
V (\vx, \vy) = g^2 \Big\langle \langle \vx | ( - \hat{\vD} \cdot \vec{\partial})^{- 1} (- \Delta) (- \hat{\vD} 
\cdot \vec{\partial})^{- 1} | \vy \rangle \Big\rangle \, .
\ee
Figure~\ref{fig-2} (b) shows the resulting quark-antiquark potential obtained from our variational Yang--Mills vacuum solution \cite{7} when the vacuum expectation
value is factorized resulting in 
\be
\label{G28}
V   (\vx, \vy) \approx \int \dd^3 w \int \dd^3 z \, G (\vx, \vw)\, \langle \vw | - \Delta | \vz \rangle 
\,G (\vz, \vy) \, ,
\ee
where $G (\vx, \vy)$ is the ghost propagator (\ref{G25}). 
The obtained potential rises linearly at large distances with a coefficient given 
by the so-called Coulomb string tension $\sigma_C$, which on the lattice is measured to be a factor of $2 \ldots 3$ 
larger than the Wilson string tension $\sigma \approx(440 \, \mbox{MeV})^2$ \cite{R4}. 
At small distances the potential behaves like the Coulomb potential as expected 
from asymptotic freedom.

\begin{figure}
\centering
\begin{subfigure}{0.45\textwidth}
\includegraphics[angle=270,width=\textwidth,clip]{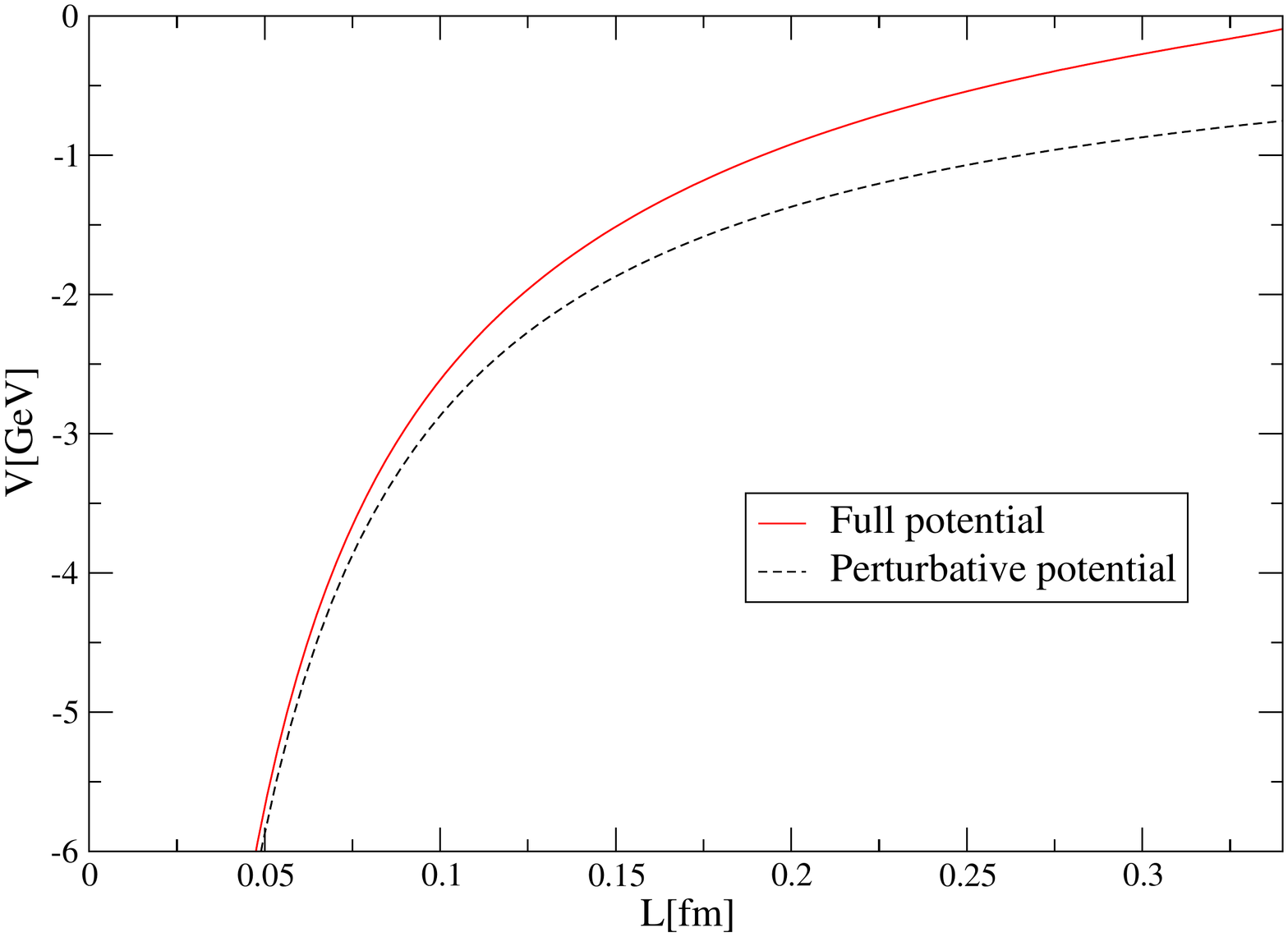}
\caption{}
\end{subfigure}
\quad
\begin{subfigure}{0.45\textwidth}
\includegraphics[angle=270,width=\textwidth,clip]{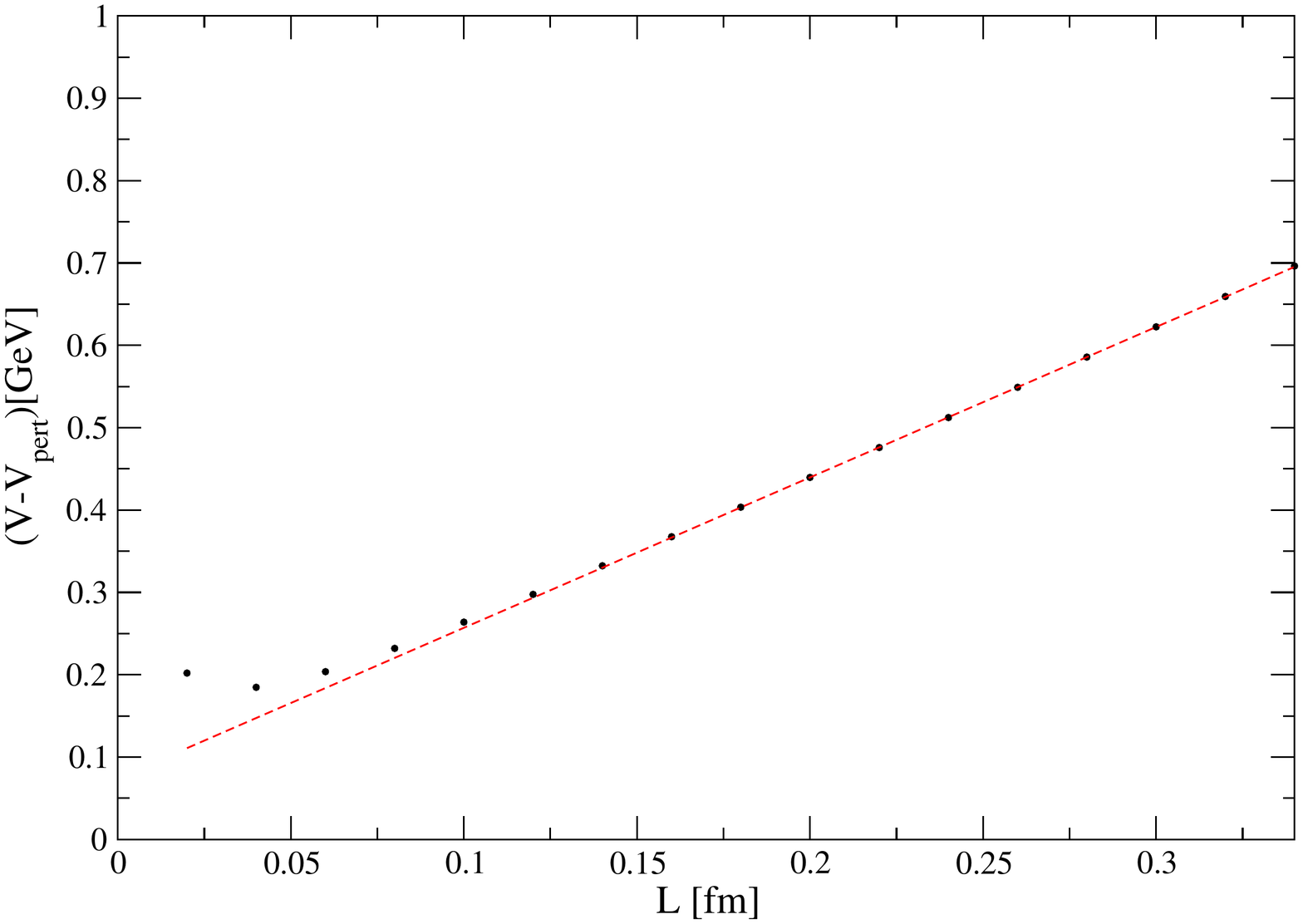}
\caption{}
\end{subfigure}
\caption{Potential obtained from the Wilson loop.}
\label{fig-6}       % Give a unique label
\end{figure}

So far,  we have seen signals for confinement both in the gluon propagator and in the static Coulomb potential. 
However, to really show that our approach yields confinement we have to calculate the vacuum expectation value 
of the Wilson loop
\be
\label{G29}
W [A] (C) = P \exp \left[ \ii \oint\limits \dd \vx \cdot \vA (\vx) \right] \, ,
\ee
which is the order parameter of confinement:
\be
\label{G30}
\langle W [A] (C) \rangle \sim
\begin{cases}
\exp (- \sigma \cA (C)) , & \text{confinement} \\[2mm]
\exp ( - \kappa \, \cP (C)) , & \text{deconfinement}
\end{cases}
\ee
In the confining theory this quantity decays exponentially with the area $\cA$ enclosed by the loop $C$ with a coefficient 
referred to as   Wilson string tension $\sigma$. An area law implies a linear rising potential with a slope given by 
$\sigma$. On the other hand in the deconfined phase the expectation value decays only with the 
perimeter $\cP$ of the loop. Unfortunately the Wilson loop is difficult to calculate in a continuum theory due to 
path ordering. In Ref.~\cite{15} an approximate evaluation of the Wilson loop was carried out and the potential 
extracted from the Wilson loop is shown in Fig.~\ref{fig-6} (a). This method works only up to 
intermediate distances where the small distance Coulomb behavior of the potential has not yet died out.
When the perturbative part is removed, however, one does indeed observe the linear rising 
potential shown in Fig.~\ref{fig-6} (b).

An order parameter (or more precisely disorder parameter) of confinement, which can be easier calculated in 
the continuum is the 't Hooft loop \cite{16}. The operator $\hat{V} (C)$ corresponding to the 't Hooft loop 
is defined by the relation 
\be
\label{G31}
\hat{V} (C_1) W (C_2) = z^{L (C_1, C_2)} W (C_2) \hat{V} (C_1) 
\ee
for any two closed loops $C_1$ and $C_2$.
Here $W (C_2)$ is the Wilson loop (\ref{G29}) and $z$ is a non-trivial center element of the gauge group. Furthermore, 
$L (C_1, C_2)$ is the Gaussian linking number, which is non-zero when the two loops in the argument are non-trivially
linked. The 't Hooft loop operator $\hat{V}(C)$ is dual to the Wilson loop in the sense that a perimeter law 
implies confinement while an area law implies deconfinement. 't Hooft gave an explicit realization of the 
loop operator $\hat{V}(C)$ on the lattice. In Ref.~\cite{17} a continuum representation of $\hat{V}(C)$ was derived,
\be
\label{G32}
\hat{V} (C) = \exp \left[\ii \il_{\R^3} \cA (C) \Pi \right] \, ,
\ee
where $\Pi = \delta /  (\ii \delta A)$ is the momentum operator of the gauge field and $\cA (C)$ is the gauge potential 
of a  so-called thin center vortex, which is a closed magnetic flux line located at the loop $C$. The center vortex field 
$\cA(C)$ is defined by its Wilson loop
\be
\label{G33}
W [ \cA (C_1) ] (C_2) = z^{L (C_1, C_2)} \, ,
\ee
which produces a non-trivial center element $z$ when the loop $C_1$ of the center vortex and the Wilson loop $C_2$ 
are non-trivially linked. Given the explicit form of the 't Hooft operator (\ref{G32}) and the coordinate representation 
of the momentum operator $\Pi = \delta / (\ii \delta A)$ it is not difficult to see that the 't Hooft loop operator,
when acting on the wave functional, displaces the argument of the wave functional by the center vortex field 
\be
\label{G34}
\hat{V} (C) \psi [A] = \psi [\cA (C) + A] \, .
\ee
With this relation the expectation value of the 't Hooft loop operator can be easily calculated by taking the overlap of the 
shifted wave functional $\psi [\cA (C) + A]$ with the unshifted one, $\psi [A]$. In Ref.~\cite{18} the 't Hooft loop was calculated 
with the wave functional obtained from our variational approach and indeed a perimeter law was found. 

\section{Variational approach to QCD in Coulomb gauge}

\begin{figure}
\centering
\includegraphics[width=0.55\textwidth,clip]{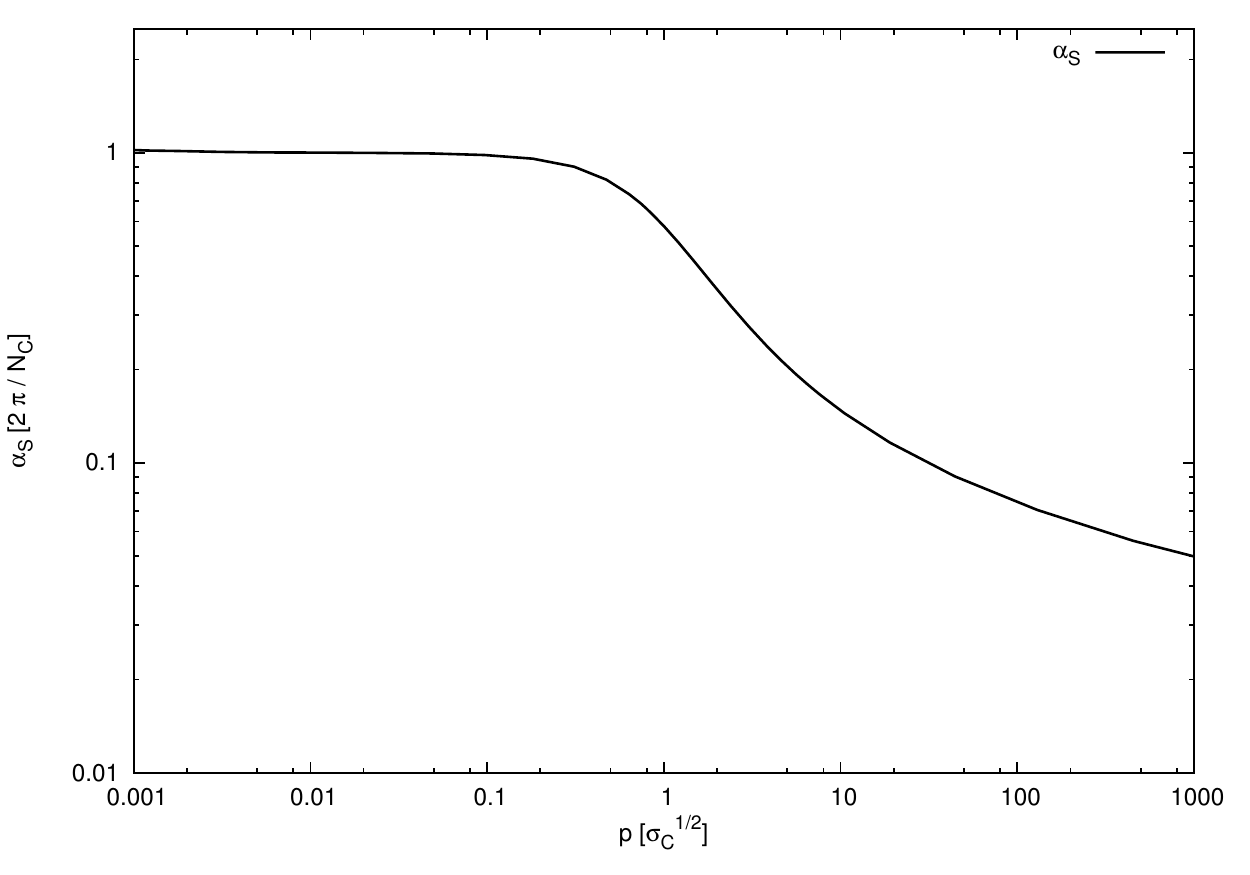}
\caption{Running coupling constant in Ref.~\cite{7} calculated in the variational approach \cite{5} from the ghost-gluon vertex \cite{25}.}
\label{fig-7}       % Give a unique label
\end{figure}

\begin{figure}
\centering
\begin{subfigure}{0.45\textwidth}
\includegraphics[trim = 60mm 0mm 60mm 0mm, width=\textwidth,clip]{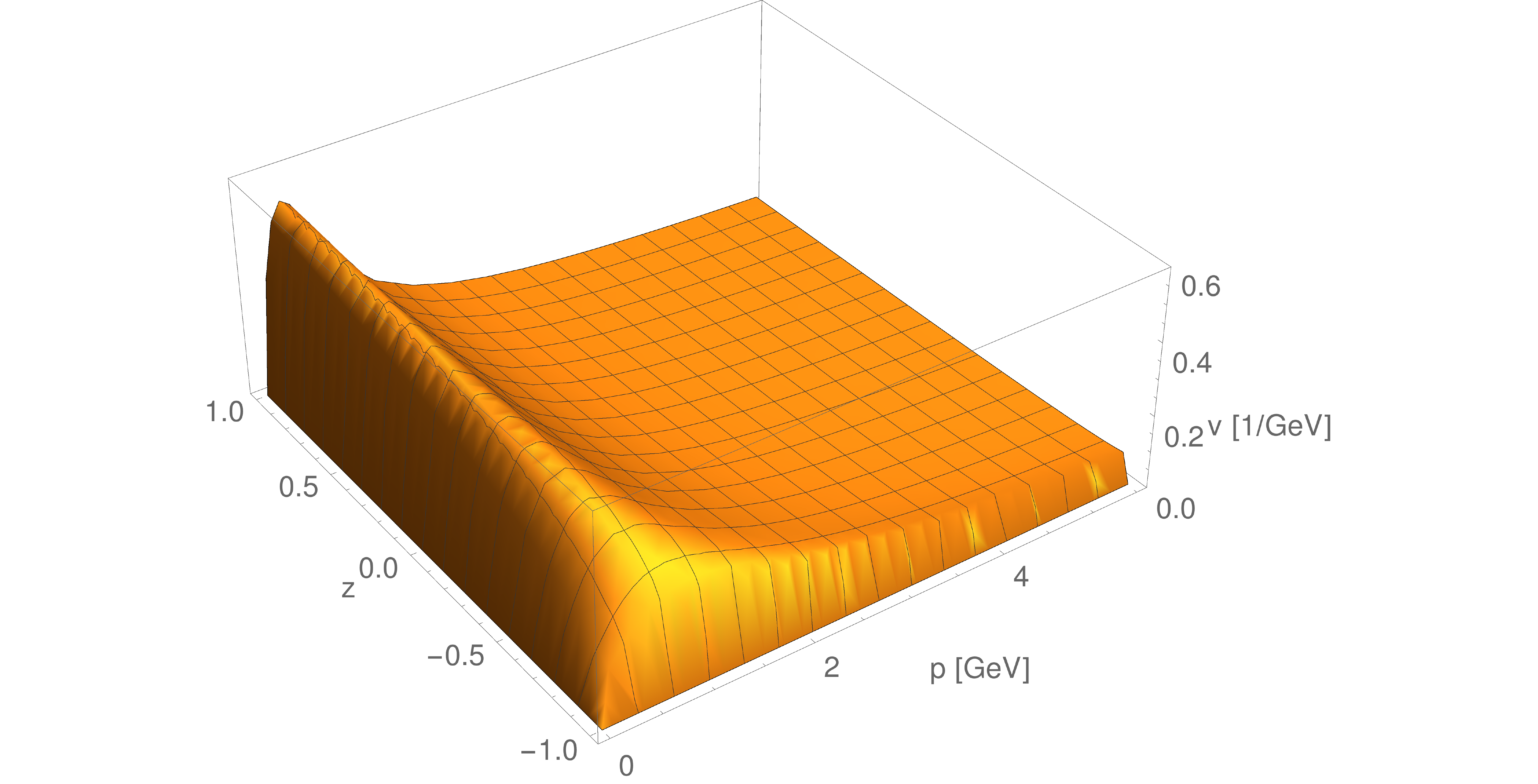}
\caption{}
\end{subfigure}
\begin{subfigure}{0.45\textwidth}
\includegraphics[trim = 60mm 0mm 60mm 0mm, width=\textwidth,clip]{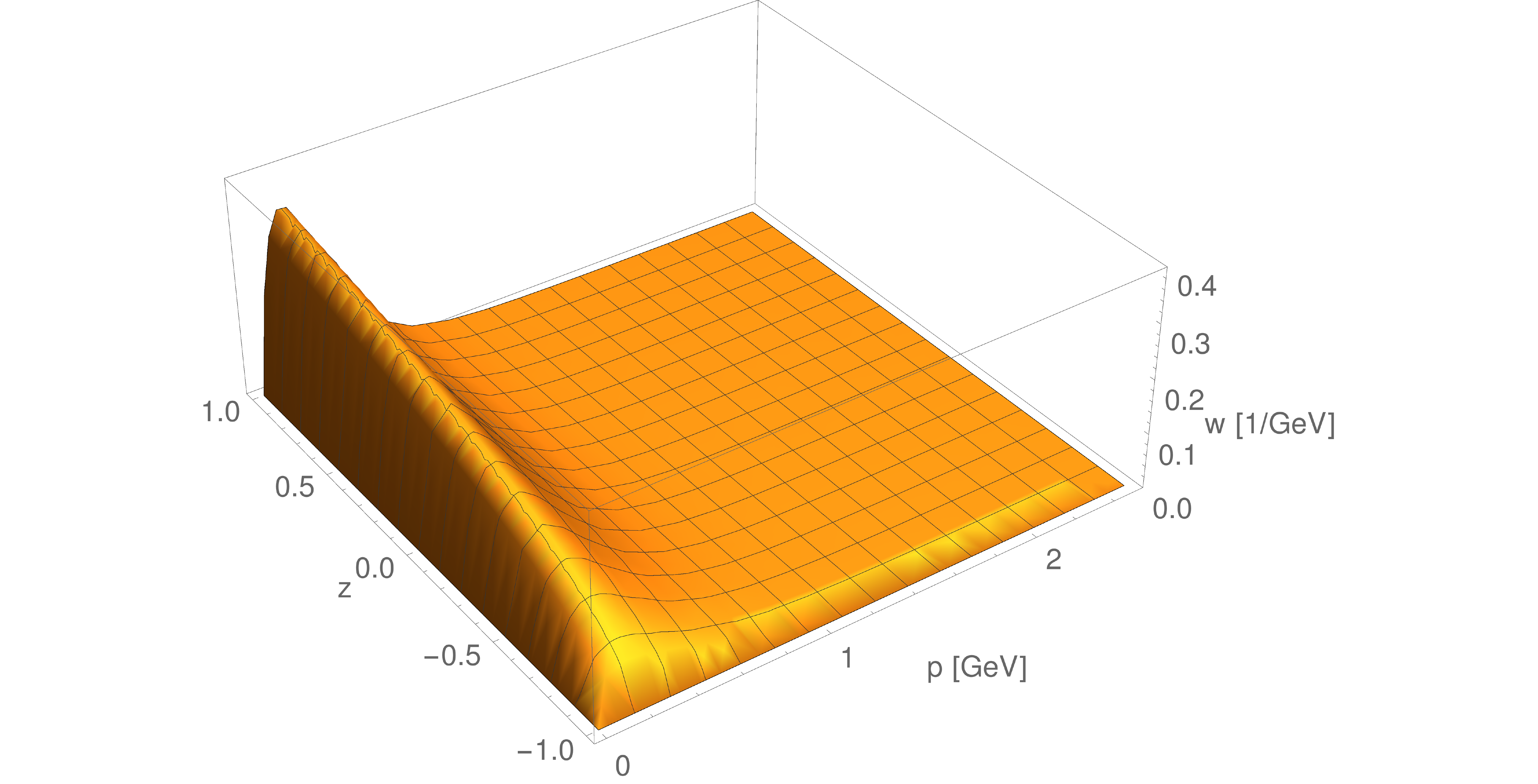}
\caption{}
\end{subfigure}
\caption{The vector kernels (a) $v (\vp, \vq)$ and (b) $w (\vp, \vq)$ obtained from the self-consistent 
solution of the variational equations of the quark sector using the Gribov formula (\ref{G26}) as input for 
the gluon energy. Notice that only the section with $p = |\vp| = |\vq|$ and $z = \vp \cdot \vq / p q$ ist plotted.}
\label{fig-8}       % Give a unique label
\end{figure}

\begin{figure}
\centering
\begin{subfigure}{0.45\textwidth}
\includegraphics[width=\textwidth,clip]{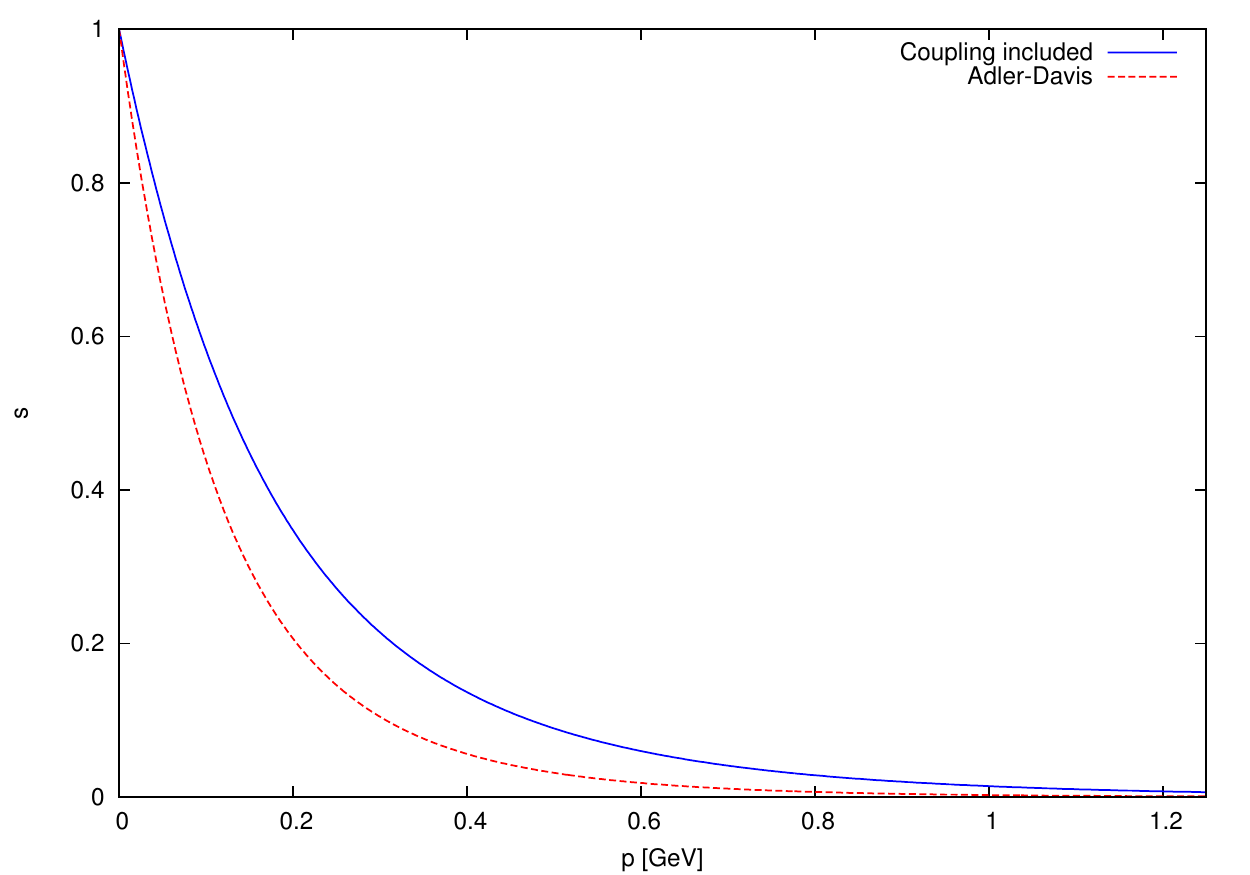}
\caption{}
\end{subfigure}
\quad
\begin{subfigure}{0.45\textwidth}
\includegraphics[width=\textwidth,clip]{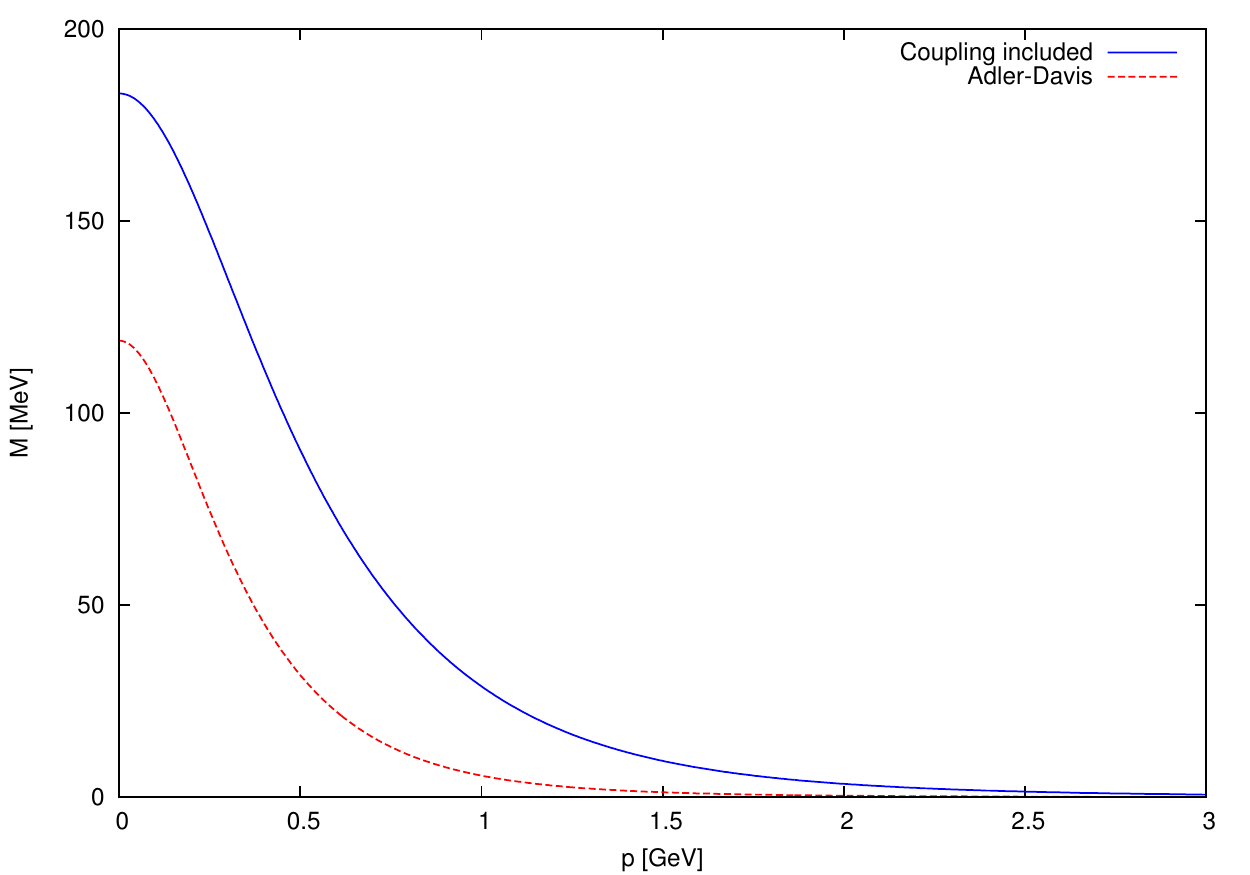}
\caption{}
\end{subfigure}
\caption{(a) The scalar kernel $s(p)$ and (b) the effective quark mass, both obtained from the self-consistent 
solution of the variational equations of the quark sector using the Gribov formula (\ref{G26}) as input for the gluon energy.}
\label{fig-9}       % Give a unique label
\end{figure}

The approach presented above for the Yang--Mills sector can be extended to full QCD. The Hamiltonian of 
QCD in Coulomb gauge is given by 
\be
\label{G35}
H_{QCD} = H_{YM} + H_C + H_q \, .
\ee
Here $H_{YM}$ is the Hamiltonian of the transversal gluon degrees of freedom given in eq.~(\ref{G10}). Furthermore, $H_C$ 
is the Coulomb term (\ref{G11}) where, however, now the charge density of the matter field is given by 
that of the quarks 
\be
\label{G36}
\rho^a_m (\vx) = \psi^\dagger (\vx) t^a \psi (\vx) \, .
\ee
Here, $\psi (\vx)$ is the quark field and $t^a$ denotes the generators of the gauge group in the fundamental
representation.  Finally, 
\be
\label{G37}
H_q = \int \dd^3 x \, \psi^\dagger (\vx) \left[ \vec{\alpha} (\vp + g \vA) + \beta m_0 \right] \psi (\vx) 
\ee
is the Dirac Hamiltonian of the quarks interacting with the spatial gauge field $\vA$. In the following let us concentrate 
on the quark sector. For the quark vacuum wave functional we have used recently the following variational ansatz \cite{19} 
\be
\label{G38}
\lvert \phi \rangle_q = \exp \left[ \int \psi^\dagger_+ (s \beta + v \vec{\alpha} \cdot \vA + w 
\beta \vec{\alpha} \cdot \vA ) \psi_- \right] \lvert 0 \rangle_q \, .
\ee
Here $\psi_\pm$ denote the positive and negative energy components of the quark Dirac spinor,  while 
$s, v, w$ are variational kernels to be determined by minimizing the QCD energy density. For $v = w = 0$ 
the ansatz reduces to a BCS type of wave functional, which was used in Refs.~\cite{20}, \cite{21}, \cite{22}. Furthermore, for 
$w = 0$ the ansatz was used in Refs.~\cite{23}, \cite{24}. There it was found that the inclusion of the coupling  of the quarks
to the transversal gluons in the vacuum wave functional substantially increases the amount of chiral symmetry breaking. 
With the quark wave functional (\ref{G38}) we have calculated the expectation value of $H_{QCD}$ up to two loops and 
carried out the variation with respect to the kernel $s$, $v$, $w$. This results in three coupled equations, where the 
equations for $v$ and $w$ can be explicitly solved in terms of the scalar kernel $s$ and the gluon propagator. 
The variational equation for the scalar kernel is usually referred to as \emph{gap equation}. 
It is highly non-linear and can only be solved numerically. The extension of the ansatz for the quark wave functional 
with $w \neq 0$ is not just a quantitative improvement (since the variational space is increased), but it also improves
the renormalization procedure, as the linear divergences in the gap equation strictly cancel.
Furthermore, the surviving logarithmic divergences always come together with the square of the coupling strength 
and can thus be absorbed into a scale dependent renormalized effective coupling constant
\be
\label{G39}
g^2 \ln \Lambda / \mu = \tilde{g}^2 (\mu) \, .
\ee
In the variational calculation of the quark sector the non-Abelian Coulomb potential shown in Fig.~\ref{fig-2} (b) enters 
as input. Through the Coulomb string tension $\sigma_C$, i.e.~the coefficient of the asymptotic linear slope of the 
potential, a scale is inherited from the gluon sector. Choosing $\sigma_C = 2 \sigma$ as suggested by lattice 
calculations \cite{R4} a coupling of 
\be
\label{G40}
\tilde{g} (\sqrt{\sigma_c}) \simeq 3.59
\ee
is needed in order to reproduce from the variational solution the phenomenological value of the 
quark condensate \cite{R7}
\be
\label{G41}
\langle \bar{q} q \rangle = ( - 235 \, \mbox{MeV})^2 \, .
\ee
In Ref.~\cite{7} the running coupling constant of Yang--Mills theory was calculated from the ghost-gluon vertex \cite{25}.
The result is shown in Fig.~\ref{fig-7}. From that figure one extracts a running coupling at the scale of the 
Coulomb string tension of 
\be
\label{G42}
\tilde{g} (\sqrt{\sigma_C}) \simeq 3.9 \, ,
\ee
which is close to the value given in eq.~(\ref{G40}) necessary to reproduce the phenomenological value of the quark
condensate. Figure~\ref{fig-8} shows the result for the variational kernels $v$ and $w$. They look quite 
similar; however, one should notice the different scales. The kernel $v$ is substantially larger than the 
kernel $w$. Figure \ref{fig-9} (a) shows the scalar form factor obtained with the coupling constant given in eq.~(\ref{G40})
and compared to the corresponding result with the BCS-type wave function where the coupling of the quarks to the spatial
gluons is neglected, i.e.~eq.~(\ref{G38}) with $v=w=0$. 
The latter  results in a quark condensate of $\langle \bar{q} q \rangle \simeq (- 165 \, \mbox{MeV})^3$,
which is substantially smaller than the phenomenological value (\ref{G41}). 
Finally, Fig.~\ref{fig-9} (b) shows the effective quark mass obtained
from the full variational calculation and the one from the BCS wave functional. Again the mass is substantially reduced when the coupling 
to the spatial gluons is  ignored. 

\section{Hamiltonian approach to finite-temperature quantum field theory by compactifying a spatial dimension}

The variational Hamiltonian approach to QCD in Coulomb gauge can be straightforwardly
extended to finite temperatures. For this purpose one uses a variational ansatz for the
density operator of the grand canonical ensemble where the exact Hamiltonian is replaced
by a single-particle Hamiltonian, with the single-particle energies determined variationally
by minimizing the thermodynamic potential or, for vanishing chemical potential, the free
energy. This has been done for the Yang--Mills sector in Refs.~\cite{8,26}.
Here I am going to present a more efficient approach to finite-temperature quantum 
field theory \cite{27,28,29}, which requires no ansatz for the density operator of
the grand canonical ensemble.

Consider finite-temperature quantum field theory in the standard functional integral approach. Here the finite temperature
is introduced by going to Euclidean space and compactifying the Euclidean time dimension by imposing periodic and 
antiperiodic boundary conditions for Bose and Fermi fields, respectively
\begin{align}
\label{656-32}
 A (x^0 = L/2) &= A (x^0 = - L / 2) \nonumber\\
 \psi (x^0 = L/2) &= - \psi (x^0 = - L/2) \, .
\end{align}
The length of the compactified dimension $L$ represents then the inverse temperature $T^{-1} = L$. One can now exploit the 
O(4) invariance of the Euclidean Lagrangian to rotate the Euclidean time axis into a space axis 
and, correspondingly, one space axis into the Euclidean  
time axis. Of course, thereby all vectorial quantities transform in the same way, i.e.~we can choose the transformation:
\begin{align}
\label{G43}
 x^0 \to x^3 \, , \quad \quad A^0 \to A^3 \, , \quad \quad \gamma^0 \to \gamma^3 \nonumber\\
 x^1 \to x^0 \, , \quad \quad A^1 \to A^0 \, , \quad \quad \gamma^1 \to \gamma^0 \, .
\end{align}
After this rotation we are left with the spatial  periodic and antiperiodic boundary conditions
\begin{align}
\label{G44}
 A (x^3 = L/2) &= A (x^3 = - L/2) \nonumber\\
 \psi (x^3 = L/2) &= - \psi (x^3 = - L/2) \, .
\end{align}
As a consequence of the O(4) rotation our spatial manifold is now $\R^2 \times S^1 (L)$ instead of $\R^3$ while 
the temporal manifold is $\R$ independent of the temperature, i.e.~the temperature is now encoded in one spatial
dimension while time has infinite extension. We can now apply the usual canonical 
Hamiltonian approach to this rotated space-time manifold. As the new time axis has infinite extension $l \to \infty$, 
the partition function is now given by 
\be
\label{G45}
Z (L) = \lim\limits_{l \to \infty} \mathrm{Tr} \exp (- l H (L)) \, ,
\ee
where $H (L)$ is the usual Hamiltonian obtained after canonical quantization, however, now defined 
on the spatial manifold $\R^2 \times S^1 (L)$. 
Taking the trace in the basis of the exact eigenstates of the Hamiltonian $H (L)$ we obtain for the partition function (\ref{G45})
\be
\label{G46}
Z (L) = \lim\limits_{l \to \infty} \sli_n \exp (- l E_n (L)) = \lim\limits_{l \to \infty} \exp (- l E_0 (L)) \, .
\ee
The full partition function is now obtained from the ground state energy calculated on the spatial manifold $\R^2 \times S^2 (L)$.
Introducing the energy density $e (L)$ on $\R^2 \times S^1 (L)$ by separating  the volume $L l^2$  of the spatial manifold
from the energy we have  
\be
\label{G47}
E_0 (L) = L l^2  e (L) \, .
\ee
It is a little exercise to show that the usual pressure $P$ is given by 
\be
\label{G48}
P = - e (L) \, ,
\ee
while the usual energy density $\varepsilon$ is obtained as 
\be
\label{G49}
\varepsilon = \partial (L e (L)) / \partial L - \mu \partial e (L) / \partial \mu \, .
\ee
To distinguish this quantity from the (negative) Casimir pressure $e(L)$ eq.~(\ref{G48}), which 
also appears as an energy density in our formalism after the transformation eq.~(\ref{G43}), we will 
denote $e(L)$ as \emph{pseudo energy density}.
Finally, after the O(4) rotation eq.~(\ref{G43}),  the finite chemical potential $\mu$ enters the 
single-particle Dirac Hamiltonian $h$ in the form 
\be
\label{G50}
h (\mu) = h(\mu=0) + \ii \mu \alpha^3 \, ,
\ee
where $\alpha^3$ is the third Dirac matrix and $h(\mu=0)$ the standard Dirac operator coupled to the gauge field. 

\begin{figure}
\centering
\begin{subfigure}{0.45\textwidth}
\includegraphics[width=\textwidth,clip]{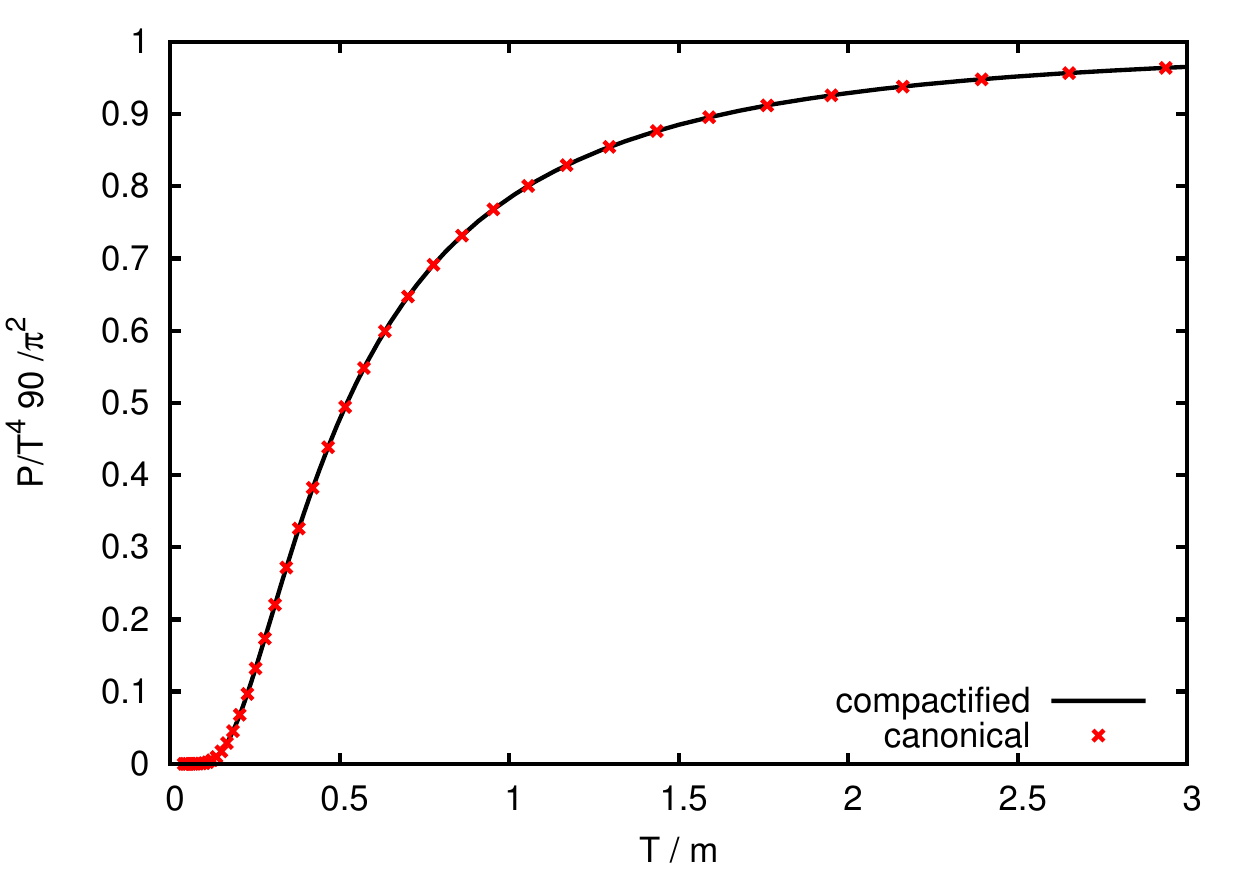}
\caption{}
\end{subfigure}
\quad
\begin{subfigure}{0.45\textwidth}
\includegraphics[width=\textwidth,clip]{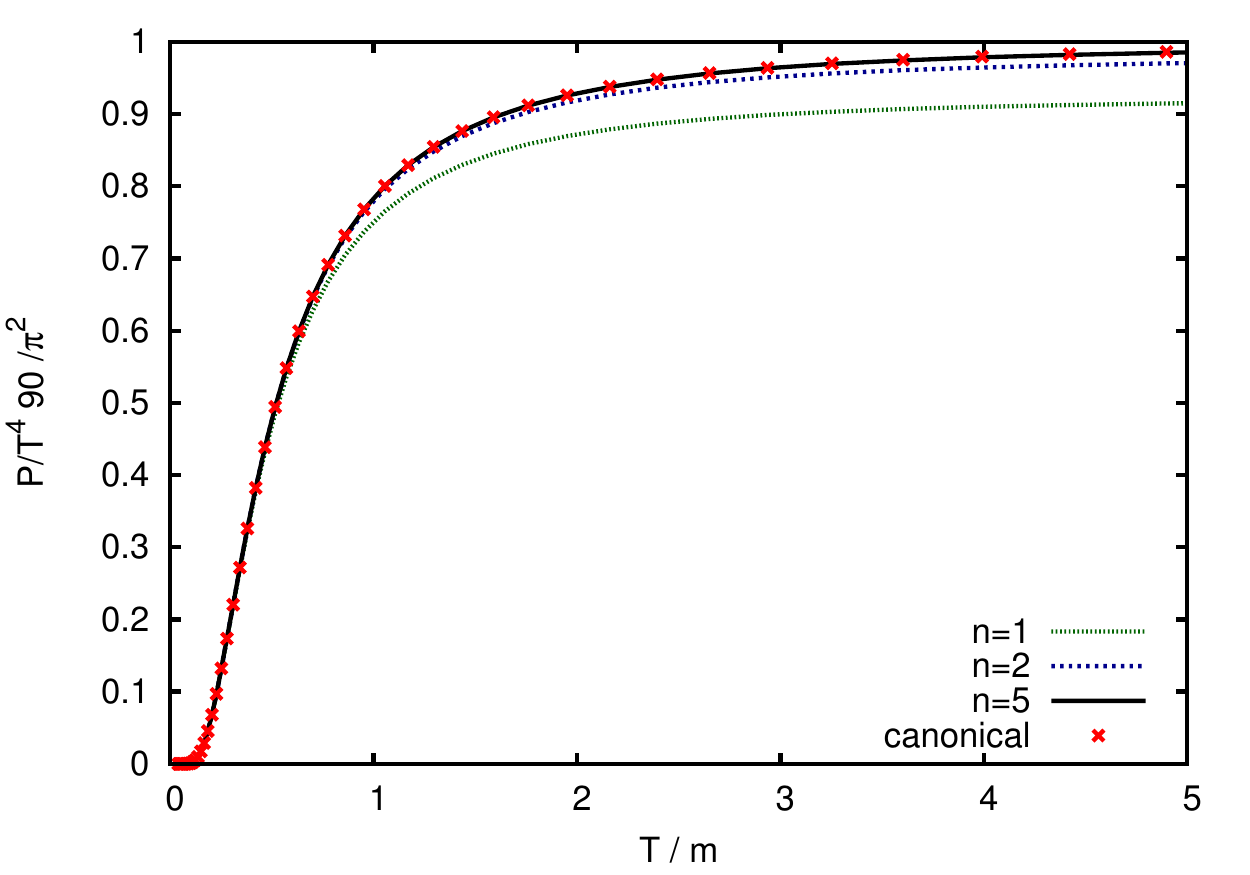}
\caption{}
\end{subfigure}
\caption{The pressure of a free massive Bose gas (a) calculated from eq.~(\ref{G55}) (full curve) and from the grand 
canonical ensemble (\ref{G51}) (crosses). (b) The pressure when the summation index in eq.~(\ref{G55}) is 
restricted to $|n| = 1, 2$ and $5$.}
\label{fig-10}       % Give a unique label
\end{figure}

To illustrate the above approach \cite{29} let us first consider a relativistic Bose gas
with dispersion relation $\omega (p) = \sqrt{\vp^2 + m^2}$, where we assume for simplicity
a vanishing chemical potential. The usual pressure obtained from the grand canonical ensemble 
for such a system is given by
\be
\label{G51}
P = \frac{2}{3} \int \frac{\dd^3 p}{(2 \pi)^3} \frac{p^2}{\omega (p)} n (p) \, , \quad \quad n (p) = \frac{1}{\exp(\beta \omega (p)) - 1} \, ,
\ee
where $n (p)$ are the finite temperature Bose occupation numbers. On the other hand for
the pseudo energy density on the spatial manifold $\R^2 \times S^1 (L)$ one finds, for the 
ideal Bose gas with dispersion relation $\omega (p) = \sqrt{\vp^2 + m^2}$ \cite{29}
\be
\label{G52}
e (L) = \frac{1}{2} \int \frac{\dd^2 p_\perp}{(2 \pi)^2} \frac{1}{L} \sli^\infty_{n = - \infty} 
\sqrt{\vp^2_\perp + {p_n}^2 + m^2} \, , 
\quad \qquad p_n = \frac{2 n \pi}{L} \, ,
\ee
where $p_n$ are the bosonic Matsubara frequencies. This quantity does not look at all like the negative of the 
pressure (\ref{G51}), as it should by eq.~(\ref{G48}). In fact, as it stands it is ill defined: 
the integral and the sum are both divergent. 
To make it mathematically well defined, we first use the proper-time regularization of the square root 
\be
\label{G53}
\sqrt{A} = \frac{1}{\Gamma \lk - \frac{1}{2} \rkx} \lim\limits_{\Lambda \to \infty}\left[ 
\il^\infty_{1/\Lambda^2} \dd \tau \,\tau^{-\frac{1}{2}}\, \exp (- \tau A) - 2 \Lambda + \mathcal{O}(\Lambda^{-1})\right] 
\, .
\ee
The divergent constant appears because the limit $\Lambda \to \infty$ of the incomplete $\Gamma$-function 
is not smooth; it drops out when taking the difference to the zero-temperature case after eq.~(\ref{G55}) below.
With this replacement, the momentum integral in eq.~(\ref{G52}) can be carried out in closed form. 
For the remaining Matsubara sum we use the Poisson resummation formula,  
\be
\label{G54}
\frac{1}{2 \pi} \sli^\infty_{k = -\infty} \exp(\ii k x) = \sli^\infty_{n = - \infty} \delta (x - 2 \pi n) ,
\ee
after which the proper-time integral can also be carried out, yielding for the pseudo energy density (\ref{G52}) 
\be
\label{G55}
e (L) = - \frac{1}{2 \pi^2} \sli^\infty_{n = - \infty} \lk \frac{m}{n L} \rkx^2 K_2
(n L m) \, ,
\ee
where $K_\nu (z)$ is the modified Bessel function. The term with $n = 0$ is divergent and represents the 
pseudo energy density at zero temperature, which should be eliminated from the pressure. The remaining terms 
$n \neq 0$ are all finite and also the remaining sum converges. This sum, however, cannot be carried out analytically for massive bosons (the same
applies to the integral in the grand canonical expression (\ref{G51}) for the pressure).
In the zero-mass limit we find from eq.~(\ref{G55}) for the pressure (\ref{G48})
\be
\label{G56}
P = \frac{\zeta(4)}{\pi^2} T^4 = \frac{\pi^2}{90} T^4 ,
\ee
which is Stefan-Boltzmann law, the correct result also obtained from the grand canonical ensemble. For massive bosons the 
evaluation of the sum in eq.~(\ref{G55}) as well as the evaluation of the integral in eq.~(\ref{G51}) have to be done 
numerically. The result is shown in Fig.~\ref{fig-10} (a). As expected the pressure calculated from the compactified spatial dimension
reproduces the result of  the usual grand canonical ensemble. Figure~\ref{fig-10} (b) shows the various contributions to the pressure.  
It is seen that only a few terms in the sum of eq.~(\ref{G55}) are necessary to reproduce the result of the grand
canonical   ensemble to good accuracy. 

In the case of the relativistic \emph{Fermi} gas with dispersion relation $\omega (p) = \sqrt{\vp^2 + m^2}$  
the energy density on $\R^2 \times S^1 (L)$  is given by 
\be
\label{G57}
e (L) = - 2 \int \frac{\dd^2 p_\perp}{(2 \pi)^2} \frac{1}{L} \sli^\infty_{n = - \infty} \sqrt{\vp^2_\perp + 
(p_n + \ii \mu)^2 + m^2} \, , \quad \quad p_n = \frac{2 n + 1}{L} \pi \, ,
\ee
where we have now included a non-vanishing chemical potential $\mu$.
To make this expression mathematically well-defined one has to resort again to  the proper-time regularization and 
Poisson resummation technique sketched above. The result is 
\be
\label{G58}
e (L) = \frac{2}{\pi^2} \sli^\infty_{n = 0} \cos \left[ n L \lk \frac{\pi}{L} - \ii \mu \rkx \right] 
\lk \frac{m}{n L} \rkx^2 K_{- 2} (n L m) \, .
\ee
Again, the term with $n = 0$ represents the zero temperature vacuum energy density, which is divergent and has to be removed.
As before, this expression can only be calculated in closed form for massless particles.
For the remaining sum to converge, an analytic continuation $\ii \mu L  \to \bar{\mu} \in \R$ is required to carry out the sum 
\be
\label{G59}
\sli^\infty_{n = 1} (-1)^n \frac{\cos (n \bar{\mu})}{n^4} = \frac{1}{48} \left[ - \frac{7}{15} \pi^2 + 2 \pi^2 \bar{\mu}^2 - \bar{\mu}^4 \right] \, .
\ee
Continuing back to real chemical potentials one finds through eq.~(\ref{G48}) for the pressure
\be
\label{G60}
P = \frac{1}{12 \pi^2} \left[ \frac{7}{15} \pi^4 T^4 + 2 \pi^2 T^2 \mu^2 + \mu^4 \right] \, ,
\ee
which is the correct result obtained also from the usual grand canonical ensemble. 

In Ref.~\cite{30} the above approach was used to study Yang--Mills theory at finite temperature. 
For this purpose, it is merely required to repeat the 
variational Hamiltonian approach on the spatial manifold $\R^2 \times S^1 (L)$.  Due to the one compactified spatial dimension 
the 3-dimensional integral equations of the zero-temperature case are replaced by a set of 2-dimensional integral 
equations distinguished by different
Matsubara frequencies. Below I will use this approach to calculate the effective potential of the Polyakov loop, the order 
parameter of confinement. 

\subsection{The Polyakov loop}

Consider SU($N$) gauge theory at finite temperature, where the temperature is introduced
by the usual periodic boundary condition in the temporal direction (\ref{656-32}). Gauge
transformations preserving this boundary conditions need to be periodic only up to an 
element $z$ of the center $Z(N)$ of the gauge group
\be
\label{G61}
U (x^0 = L) = z U (x^0 = 0) \, , \quad \quad z \in Z (N) \, .
\ee
Since there are $N$ center elements this theory has a residual global $Z (N)$ symmetry, 
which remains after gauge fixing. However, there are quantities which are sensitive to such a $Z(N)$ 
symmetry transformation. The most prominent  example is the Polyakov loop 
\be
\label{G62}
P [A_0] (\vx) = \frac{1}{d} \mathrm{tr} P \exp \left[ \ii \il^L_0 \dd x^0 \, A_0 (\vx, x^0) \right] \, ,
\ee
where $A_0 = A^a_0\,t^a$ is the temporal gauge field in the fundamental representation, and $d$ is the dimension 
of this representation. A gauge transformation of the form (\ref{G61}) multiplies the Polyakov loop by the 
center element $z$
\be
\label{G63}
P [A^U_0] = z P [A_0] \, .
\ee
The expectation value of the Polyakov loop 
\be
\label{G64}
\langle P [A_0] (\vx) \rangle \sim \exp \lk - F_\infty (\vx) L \rkx
\ee
can be shown to be related to the free energy $F_\infty (\vx)$ of a static color point charge located at 
$\vx$ \cite{R5}. 
In a confining theory this quantity has to be infinite since there are no free color charges, while in a 
deconfined phase it is finite. Accordingly we find for the expectation value of 
the Polyakov loop 
\be
\label{G65}
\langle P [A_0] (\vx) \rangle \left\{ 
\begin{array}{ll}
 =  0 & \mbox{confined phase,} \\
 \neq 0 & \mbox{deconfined phase.} 
 \end{array}
\right.
\ee
From eq.~(\ref{G63}) follows that a state with vanishing expectation value of the Polyakov loop is obviously invariant with respect to the 
global center transformation, while in the deconfined phase the $Z (N)$ center symmetry is obviously broken.
In the continuum theory the Polyakov loop can be most easily calculated in the Polyakov gauge
\be
\label{G66}
\partial_0 A_0 = 0 , \qquad A_0 \text{ color diagonal.}
\ee
In this gauge one finds, for example, for the SU(2) gauge group that  the Polyakov loop
\be
\label{G67}
P [A_0] (\vx) = \cos \lk \frac{1}{2} A_0 (\vx) L \rkx \, 
\ee
is a unique function of the gauge field, at least in the fundamental modular region of this gauge. It can be shown, see 
Refs.~\cite{31}, \cite{32}, that instead of the expectation value of the Polyakov loop $\langle P [A_0] \rangle$ one 
may alternatively use the Polyakov loop of the expectation value, $P [\langle A_0 \rangle ]$, or the expectation value 
of the temporal gauge field itself, $\langle A_0 \rangle$, as order parameter of confinement. This analysis also 
shows that the most efficient way to obtain the Polyakov loop is to carry out a so-called background field calculation 
with a temporal background field $a_0 (\vx) = \langle A_0 (\vx) \rangle$ chosen in the Polyakov gauge, and then 
calculate the effective potential $e [a_0]$ of that background field. From the minimum $\bar{a}_0$ of this potential 
one evaluates the Polyakov loop $P[\langle A_0\rangle] = P [\bar{a}_0]$, which can then serve as the order parameter of confinement. 

\begin{figure}
\centering
\begin{subfigure}{0.45\textwidth}
\includegraphics[width=\textwidth,clip]{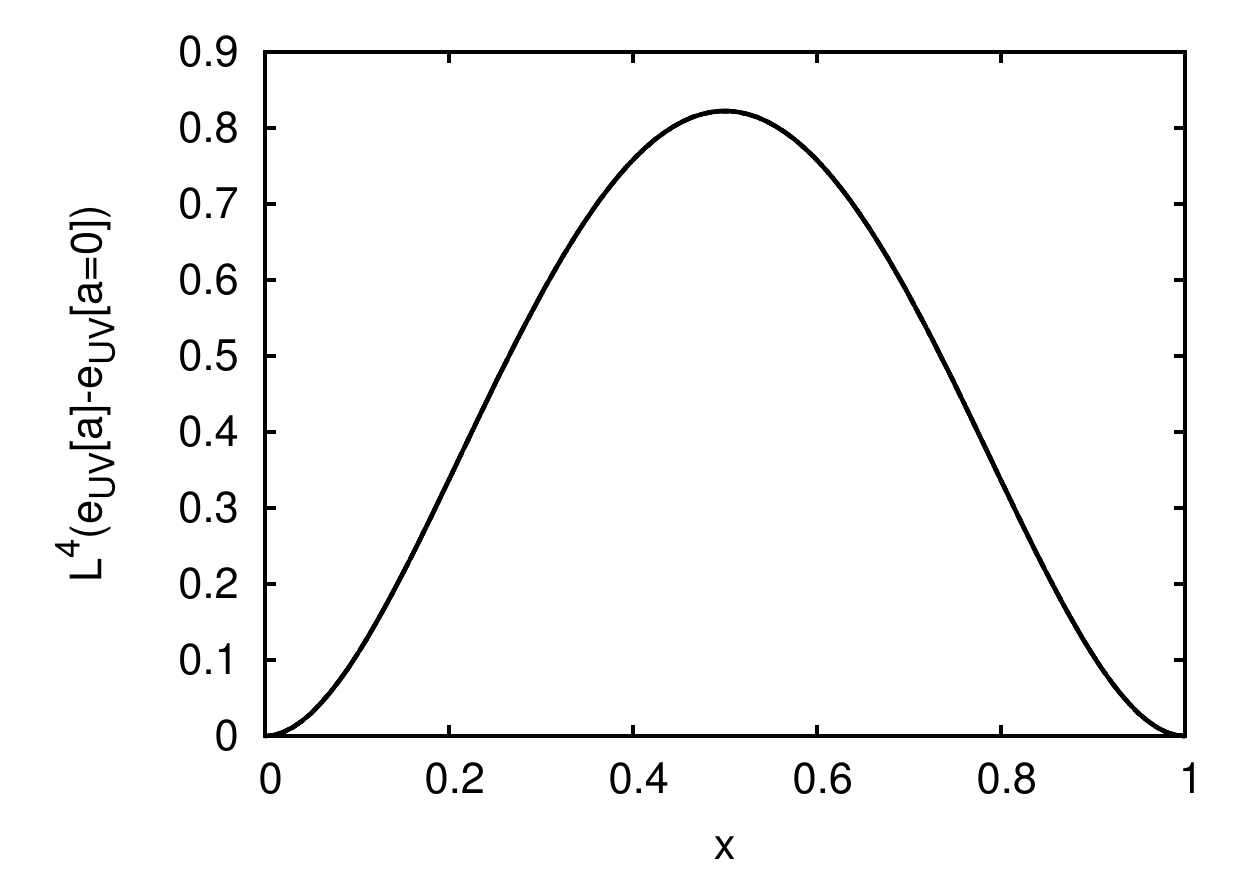}
\caption{}
\end{subfigure}
\quad
\begin{subfigure}{0.45\textwidth}
\includegraphics[width=\textwidth,clip]{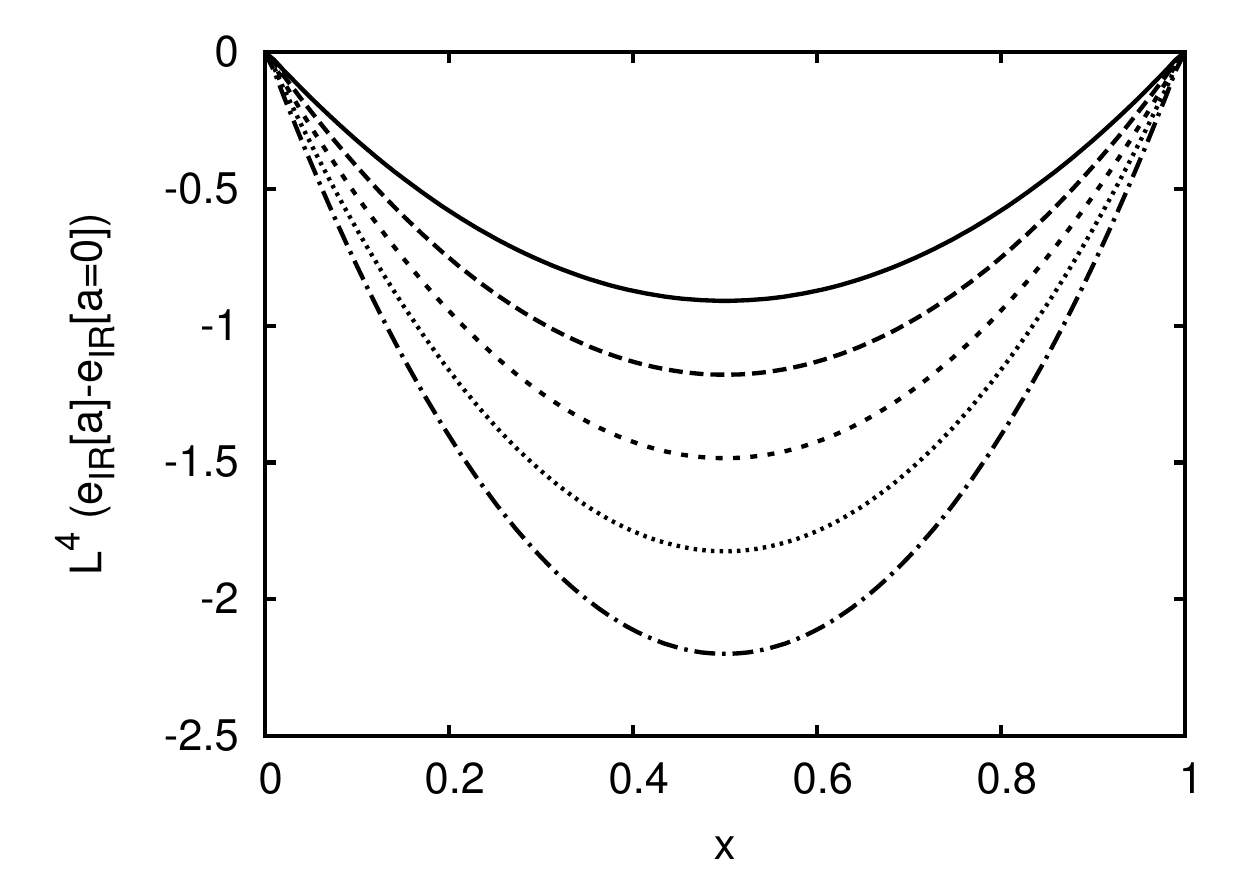}
\caption{}
\end{subfigure}
\caption{The effective potential of the Polyakov loop $e(a, L)$ (\ref{G68}) as function of the 
background field $x = a_3 L / 2 \pi$. The curvature is neglected $(\chi = 0)$ and the gluon energy assumed 
to be (a) $\omega (p) = p$ (UV-form) and (b) $\omega (p) = M^2 / p$ (IR-form) respectively.}
\label{fig-11}       % Give a unique label
\end{figure}

Such a calculation was done a long time ago in Ref.~\cite{33}, where the effective potential $e [a_0]$ was calculated 
in 1-loop perturbation theory. The result is shown in Fig.~\ref{fig-11} (a). The potential is periodic due to center symmetry. 
The minimum of the potential occurs at the vanishing background field, which gives $P [a_0 = 0] = 1$ corresponding 
to the deconfined phase. This is, of course, expected due to the use of perturbation theory.  Below I present the results of a 
non-perturbative evaluation of $e [a_0]$ in the Hamiltonian approach in Coulomb gauge. 

At first sight it seems that the Polyakov loop cannot be calculated in the Hamiltonian approach due to the use of the Weyl gauge $A_0 = 0$. However, we can 
now use the alternative Hamiltonian approach to finite temperature introduced above, where the temperature is introduced by compactifying a spatial
dimension. Here, we compactify the $x_3-$axis and consequently put also the background field along this axis $\va = a \ve_3$. In the Hamiltonian 
approach the effective potential of a spatial background field $\va$  can be easily calculated by minimizing the expectation value of the Hamiltonian
under the constraint $\langle \vA \rangle = \va$. The resulting energy 
$\langle H \rangle_{\va} = L^2 l e (\va)$ is then (up to the spatial volume factor) the effective potential. 
So the effective potential $e (\va)$ is nothing but the pseudo energy density considered earlier, but now 
calculated in a background gauge with the contraint $\langle \vA \rangle = \va$. 

\subsection{The effective potential of the Polyakov loop}

After lengthy calculations exploiting the gluon gap equation (\ref{G14}) one finds for the effective potential of the Polyakov loop the following expression
\be
\label{G68}
e (a, L) = \sli_\sigma \frac{1}{L} \sli^\infty_{n = - \infty} \int \frac{\dd^2 p_\perp}{(2 \pi)^2} \lk \omega (\vp^\sigma) - \chi (\vp^\sigma) \rkx \, .
\ee
Here $\omega (p)$ is the gluon energy and $\chi (p)$ is the ghost loop. Furthermore, these quantities have to be taken with the momentum variable
\be
\label{G69}
\vp^\sigma = \vp_\perp + \lk p_n - \sigma \cdot a \rkx \ve_3 \, ,
\ee
where $\vp_\perp$  is the momentum corresponding to the two uncompactified space dimensions while $p_n = 2 \pi n / L$ is the Matsubara frequency resulting 
from the compactification of the third dimension. Furthermore, $\sigma \cdot a \equiv \sigma^b a^b$ denotes the product of the color background field with the root vectors 
$\sigma^b$ of the gauge group. Equation~(\ref{G68}) includes also the summation over the roots $\sigma$ of the gauge group. In Ref.~\cite{27,28} the effective potential
(\ref{G68}) was explicitly calculated using for $\omega (p)$ and $\chi (p)$ the results from the variational calculation in Coulomb gauge at zero temperature \cite{7}. This
represents certainly an approximation since, in principle, one should use the finite-temperature solutions obtained in Ref.~\cite{30}. 

\begin{figure}
\centering
\begin{subfigure}{0.45\textwidth}
\includegraphics[width=\textwidth,clip]{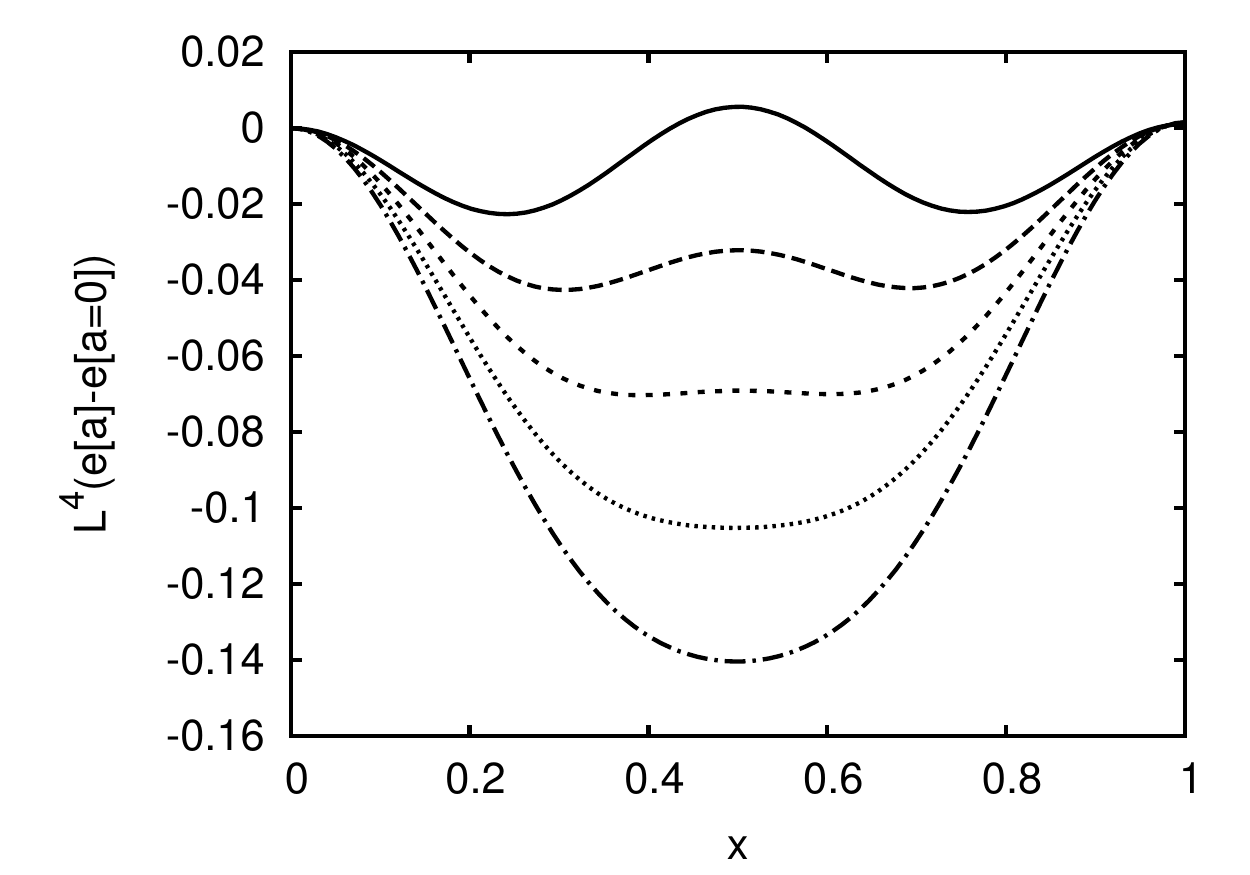}
\caption{}
\end{subfigure}
\quad
\begin{subfigure}{0.45\textwidth}
\includegraphics[width=\textwidth,clip]{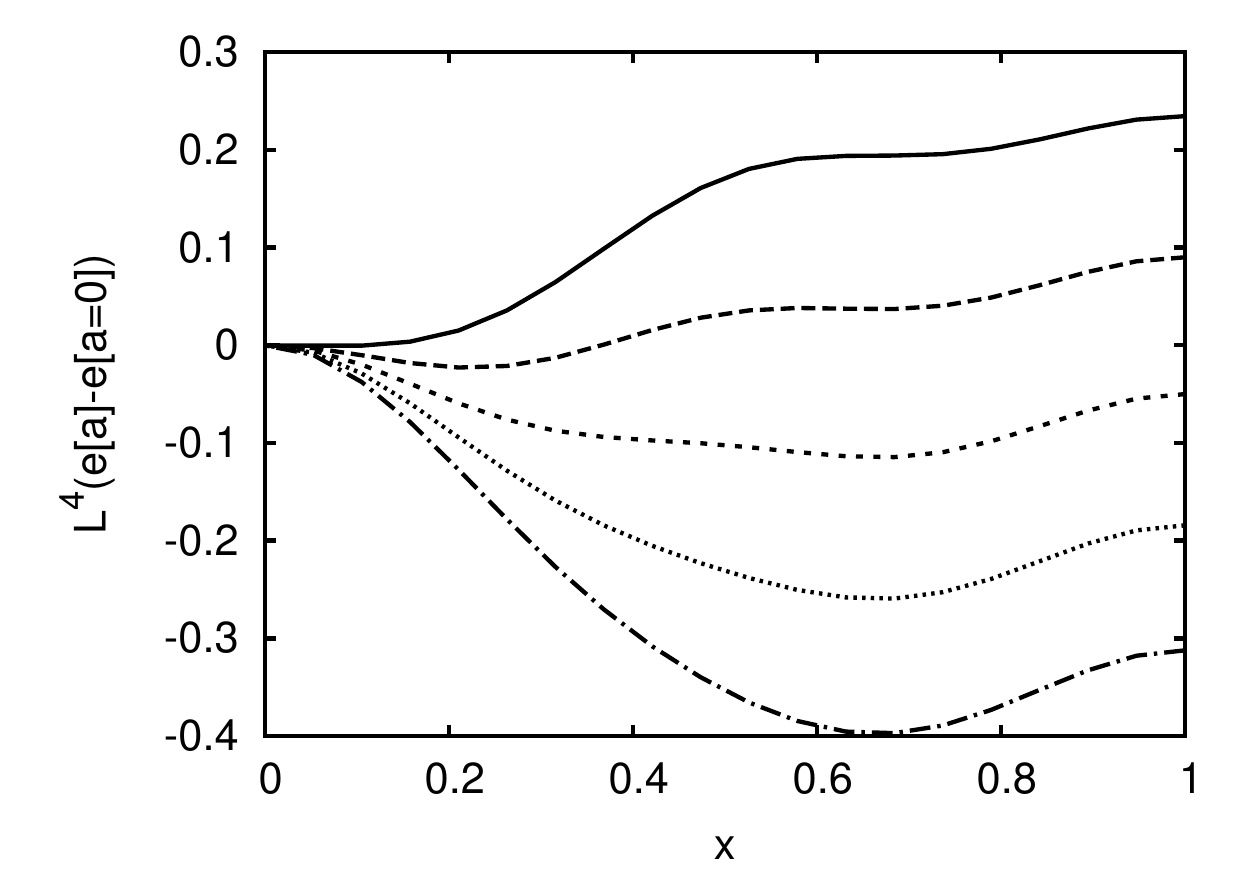}
\caption{}
\end{subfigure}
\caption{Effective potential of the Polyakov loop as function of the background field $x = a_3 L / 2 \pi$ 
at various temperatures, for the gauge group (a) SU(2) and (b) SU(3).}
\label{fig-12}       % Give a unique label
\end{figure}

\begin{figure}
\centering
\begin{subfigure}{0.45\textwidth}
\includegraphics[width=\textwidth,clip]{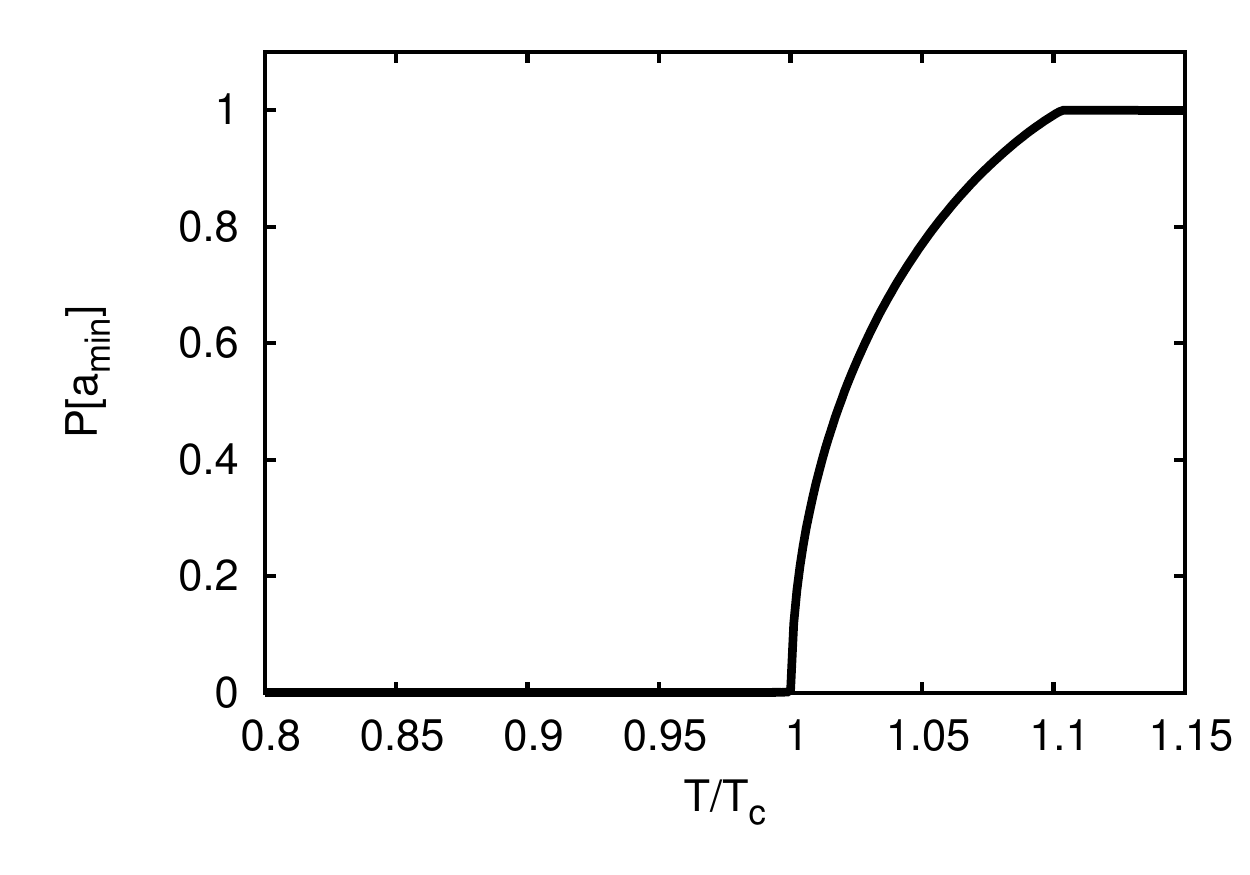}
\caption{}
\end{subfigure}
\quad
\begin{subfigure}{0.45\textwidth}
\includegraphics[width=\textwidth,clip]{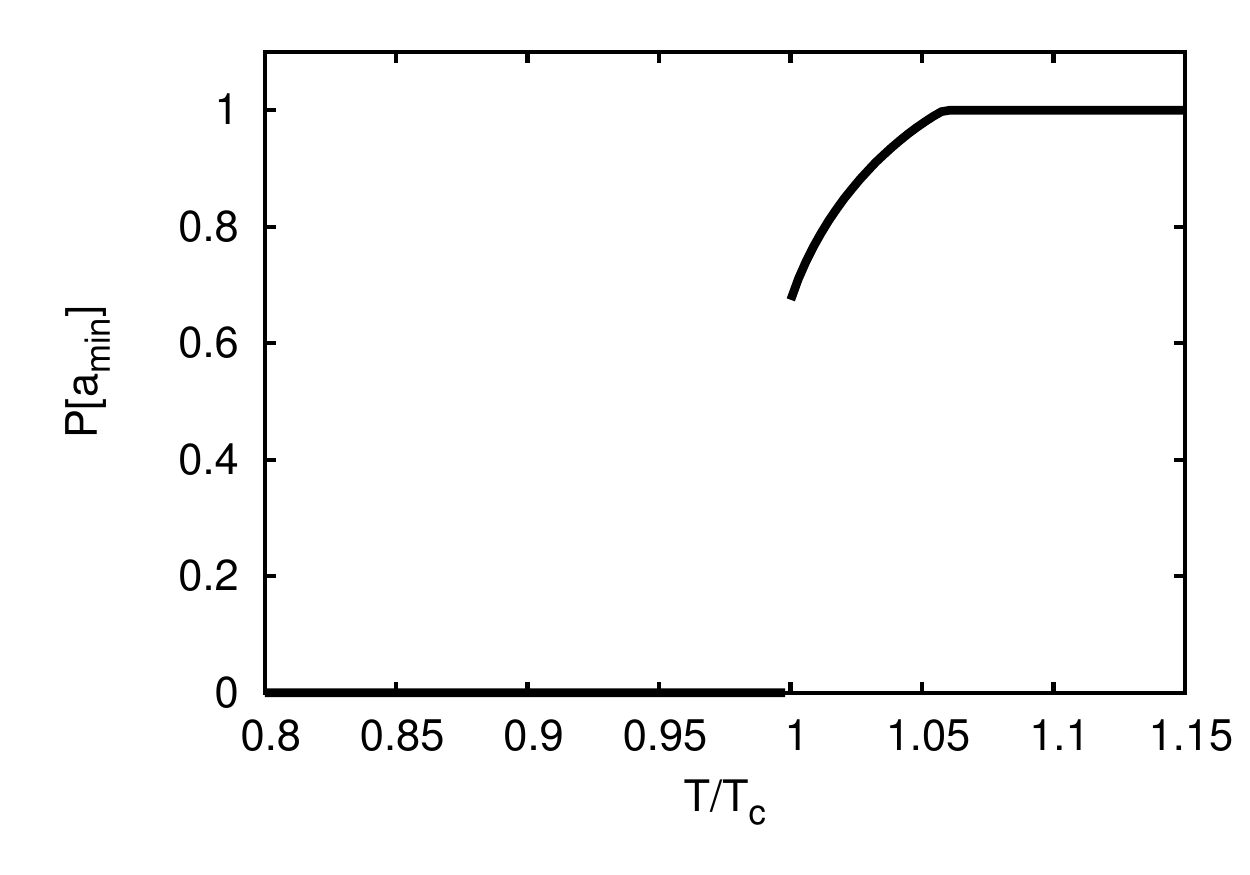}
\caption{}
\end{subfigure}
\caption{The Polyakov loop as function of the temperature (a) for SU(2) and (b) for SU(3).}
\label{fig-13}       % Give a unique label
\end{figure}

Before I present the full results let me ignore the ghost loop $\chi (p)$ in eq.~(\ref{G68}) and consider the ultraviolet and infrared limit of the gluon energy. If we 
choose the ultraviolet limit $\omega (p) = p$, we obtain from eq.~(\ref{G68}) with $\chi (p) = 0$ precisely the Weiss potential, shown in Fig.~\ref{fig-11} (a), which 
corresponds to the deconfined phase. Choosing for the gluon energy its infrared limit $\omega (p) = M^2 / p$ one finds from eq.~(\ref{G68}) with 
$\chi (p) = 0$ the (center symmetric) potential shown in Fig.~\ref{fig-11} (b). From its center symmetric minimum $\bar{a} = \pi / L$ one finds a vanishing Polyakov
loop $P [\bar{a}] = 0$ corresponding to the confined phase. Obviously, the deconfining phase transition results from the interplay between the 
confining infrared and the deconfining ultraviolet potentials. Choosing for the gluon energy the sum of the UV- and IR-parts $\omega (p) = p + M^2/p$,
which can be considered as an approximation to the Gribov formula (\ref{G26}), one has to add the UV and IR potentials and finds a phase transition
at a critical temperature $T_c = \sqrt{3} M / \pi$. With the Gribov mass $M \approx 880 \,$ MeV this gives a critical value of 
$T_c \approx 485 \,$ MeV for the color group SU(2),
which is much too high as compared to the lattice value of $312\,\text{MeV}$ \cite{R6}. 
One can show analytically \cite{27,28} that the neglect of the ghost loop $\chi (p) = 0$ shifts the critical temperature to higher
values. If one uses for the gluon energy $\omega (p)$ the Gribov formula (\ref{G26}) and includes the ghost loop $\chi (p)$ one finds the effective 
potential shown in Fig.~\ref{fig-12} (a), which shows a second order phase transition and gives a transition temperature of $T_c \approx 269 \, $ MeV for the gauge group SU(2), which is in the right 
ballpark. The Polyakov loop $P [\bar{a}]$ calculated from the minimum $\bar{a}$ of the effective potential $e (a, L)$ (\ref{G68}) is plotted in Fig.~\ref{fig-13}(a)
as function of the temperature.
\begin{figure}
\centering
\begin{subfigure}{0.45\textwidth}
\includegraphics[width=\textwidth,clip]{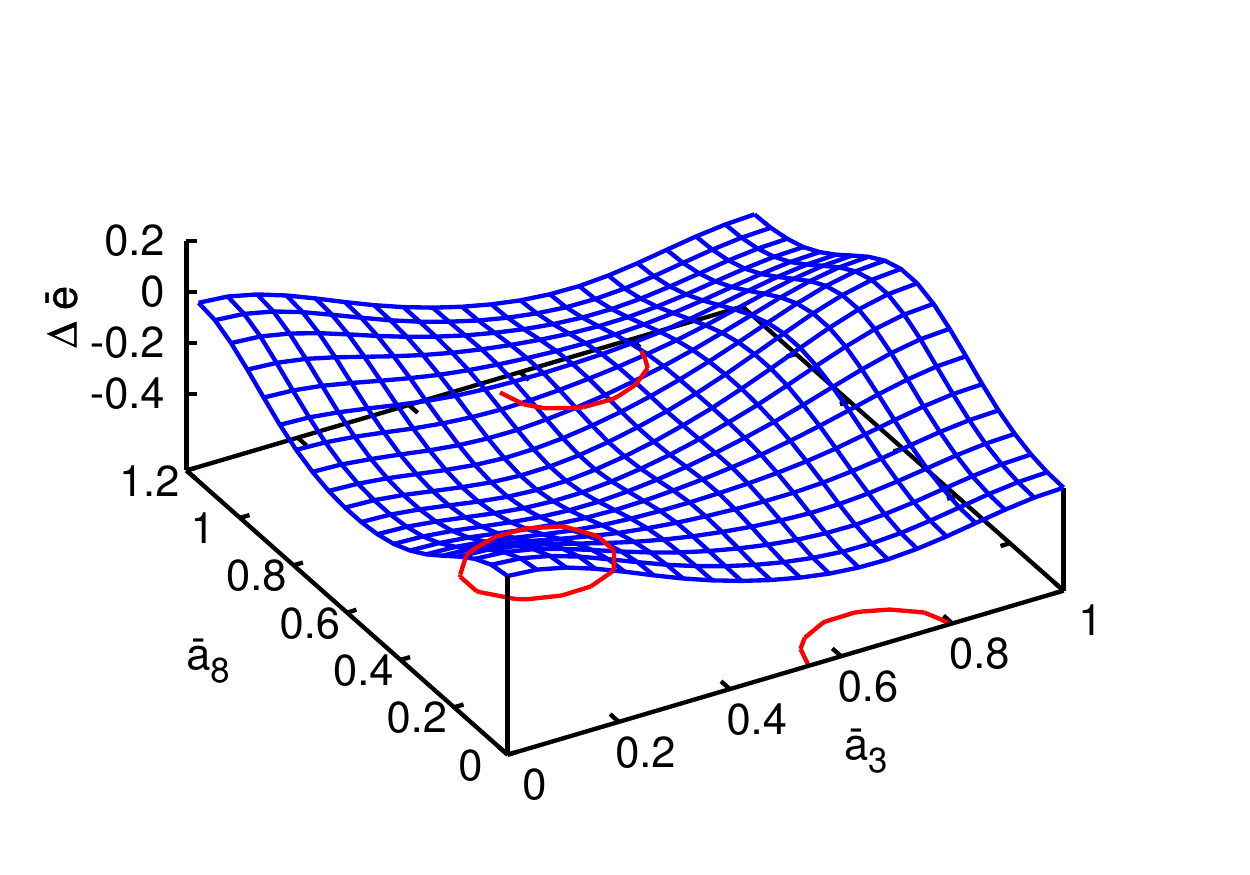}
\caption{}
\end{subfigure}
\quad
\begin{subfigure}{0.45\textwidth}
\includegraphics[width=\textwidth,clip]{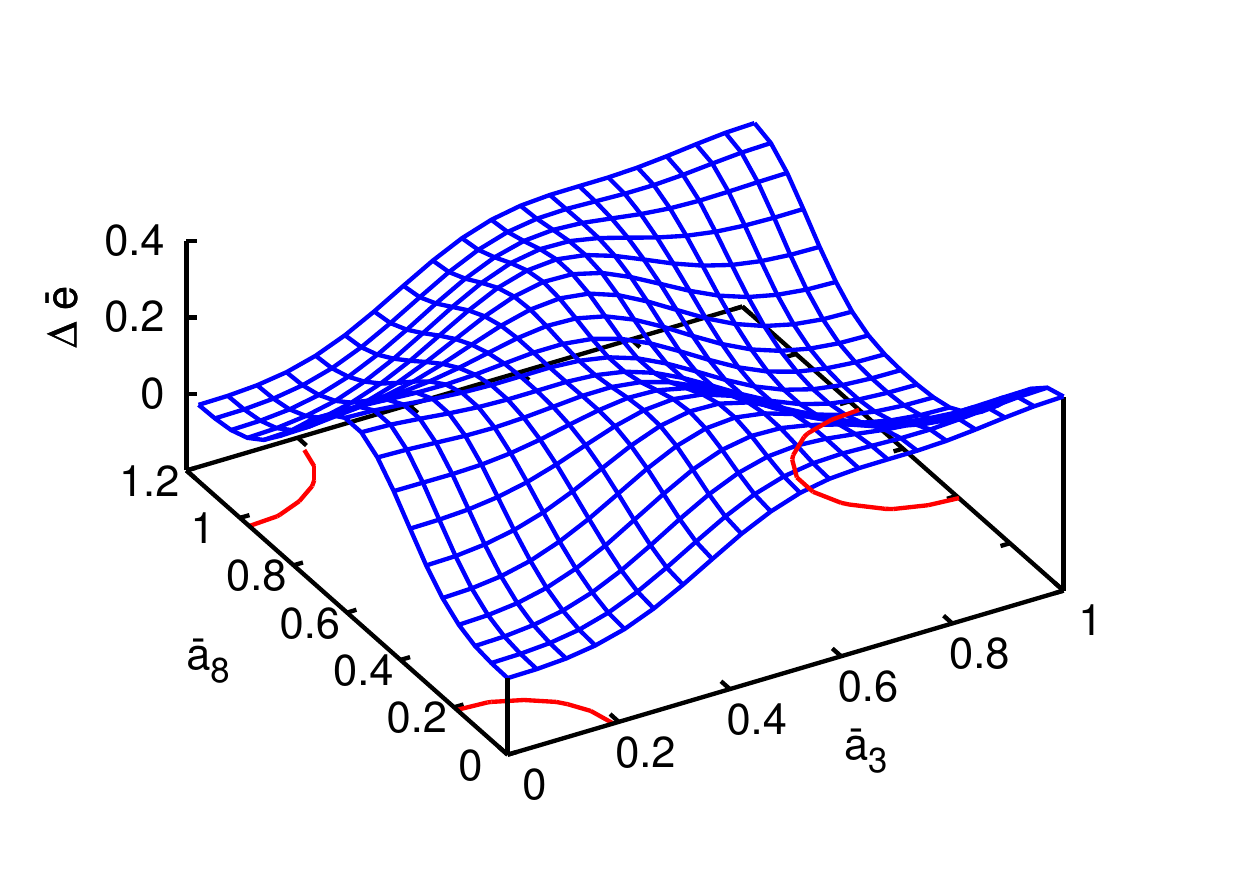}
\caption{}
\end{subfigure}
\caption{The effective potential of the Polyakov loop for the gauge group SU(3) as function of the two Cartan 
components of the background field $x = a_3 L / 2 \pi$ and $y = a_8 L / 2 \pi$ for (a) $T < T_c$ and (b) $T > T_c$.}
\label{fig-14}       % Give a unique label
\end{figure}

The effective potential for the gauge group SU(3) can be reduced to that of the SU(2) group by noticing that the SU(3) algebra consists of 
three SU(2) subalgebras characterized by the three positive roots $\vec{\sigma} = (1, 0) \, , \, (1/2, 1/2\sqrt{3}) \, , \, (1/2 \, , - 1 /2 \sqrt{3})$.
One finds 
\be
\label{931-70}
e_{\mathrm{SU(3)}} (a, L) = \sum_{\boldsymbol{\sigma} > 0} e_{\mathrm{SU(2)}} [\vec{\sigma}] (a, L) \, .
\ee
The resulting effective potential for SU(3) is shown in Fig.~\ref{fig-14} as function of the components of the background field in the Cartan algebra $a_3, a_8$.
Above and below $T_c$ the absolute minima of the potential occur in both cases for $a_8 = 0$. Cutting the 2-dimensional potential surface at $a_8 = 0$ 
one finds the effective potential shown in Fig.~\ref{fig-12} (b), which shows a first order phase transition with a critical temperature of $T_c \approx 283 \,$ MeV. 
The first order nature of the SU(3) phase transition is also seen in Fig.~\ref{fig-13} (b), where the Polyakov loop $P [\bar{a}]$ is shown as function of the 
temperature. 

\section{Conclusions}

In this talk I have shown that the Hamiltonian approach to QCD in Coulomb gauge gives a decent description of the infrared 
properties of the theory and at the same time can be straightforwardly extended to finite temperatures, where it yields 
critical temperatures for the deconfinement phase transition in the right ballpark. Furthermore, the approach has also 
produced the correct order of the deconfinement phase transition, viz.~first order for SU(3) and second order for SU(2). 
The obtained results are very encouraging and call for an extension of the present approach to non-zero chemical potentials.

\end{document}